\documentclass[aps,pr,twocolumn,showpacs,superscriptaddress,groupedaddress,nofootinbib]{revtex4}  

\usepackage{longtable}
\usepackage{graphicx}  
\usepackage{ulem}   
\usepackage{comment} 
\usepackage{datetime}
\usepackage{dcolumn}

\usepackage{amsxtra}
\usepackage{euscript} 
\usepackage{bm}        
\usepackage{tabularx}
\usepackage{float}
\restylefloat{table}

\usepackage[cmtip,arrow]{xy}
\usepackage{pb-diagram,pb-xy}

\usepackage{amsmath}
\usepackage{amssymb}
\usepackage{amsthm}
\usepackage[pdftex]{color}
\usepackage[pdftex,colorlinks,citecolor=blue,linkcolor=blue,urlcolor=blue]{hyperref} 
\usepackage{graphicx}
\usepackage{dcolumn} 
\usepackage{bm} 

\usepackage{longtable}

\usepackage{ulem}   
\usepackage{comment} 
\usepackage{datetime}

\usepackage{tabularx}
\usepackage{float}
\restylefloat{table}

\newcommand{\beq}{\begin{equation}}
\newcommand{\eeq}{\end{equation}}
\newcommand{\bea}{\begin{eqnarray}}
\newcommand{\eea}{\end{eqnarray}}
\newcommand{\barr}{\begin{array}}
\newcommand{\earr}{\end{array}}

\long\def\begincomment#1\endcomment{}

\newcommand{\U}{\mathrm{U}}

\newcommand{\Tr}{\mathrm{Tr}}
\newcommand{\perm}{\mathrm{perm}}
\newcommand{\Sym}{\mathrm{Sym}}

\newtheorem{definition}{Definition}

\newtheorem{proposition}{Proposition}
\newtheorem{claim}{Claim}
\usepackage{graphicx}

\usepackage{pgf,tikz}
\usetikzlibrary{arrows}

\usepackage{bm}
\usepackage{bbold}

\usepackage{mathtools}

\DeclarePairedDelimiterX\braket[2]{\langle}{\rangle}{#1 \delimsize\vert #2}

\begin{document}


\title{Reliability of the local truncations for the random tensor models renormalization group flow}

\author{Vincent Lahoche} \email{vincent.lahoche@cea.fr}   
\affiliation{Commissariat à l'\'Energie Atomique (CEA, LIST),
 8 Avenue de la Vauve, 91120 Palaiseau, France}
\author{Dine Ousmane Samary}
\email{dine.ousmanesamary@cipma.uac.bj}
\affiliation{Commissariat à l'\'Energie Atomique (CEA, LIST),
 8 Avenue de la Vauve, 91120 Palaiseau, France}
\affiliation{International Chair in Mathematical Physics and Applications (ICMPA-UNESCO Chair), University of Abomey-Calavi,
072B.P.50, Cotonou, Republic of Benin}

\begin{abstract}
\begin{center}
\textbf{Abstract}
\end{center}
The standard nonperturbative approaches of renormalization group  for tensor models are generally focused on a purely local potential approximation (i.e. involving only generalized traces and product of them) and are showed to strongly violate the modified Ward identities. This paper as a continuation of our recent contribution [Physical Review D 101, 106015 (2020)], intended to investigate the approximation schemes compatibles with Ward identities and constraints between $2n$-points observables in the large $N$-limit. We consider separately two different approximations: In the first one, we try to construct a local potential approximation from a slight modification of the Litim regulator, so that it remains optimal in the usual sense, and preserves the boundary conditions in deep UV and deep IR limits. In the second one, we introduce derivative couplings in the truncations and show that the compatibility with Ward identities implies strong relations between $\beta$-functions, allowing to close the infinite hierarchy of flow equations in the non-branching sector, up to a given order in the derivative expansion. Finally, using exact relation between correlations functions in large $N$-limit, we show that strictly local truncations are insufficient to reach the exact value for the critical exponent, highlighting the role played by these strong relations between observables taking into account the behavior of the flow; and the role played by the multi-trace operators, discussed in the two different approximation schemes. In both cases, we compare our conclusions to the results obtained in the literature and conclude that, at a given order, taking into account the exact functional relations between observables like Ward identities in a systematic way we can strongly improve the physical relevance of the approximation for exact RG equation. \\

\noindent
\textbf{Key words :} Random tensor models, discrete gravity, quantum gravity, random geometry, renormalization group.
\end{abstract}

\pacs{11.10.Gh, 02.40.Gh, 11.10.Hi, 04.60.-m}

\maketitle

\section{Introduction}\label{section0}

Random tensor models (RTMs) were initially introduced in the quantum gravity context at the beginning of 1990's \cite{Ooguri:1991ib}-\cite{Gross:1991hx}, as a natural extension of random matrix models (RMMs) used to quantize $2$-dimensional gravity \cite{Brezin:1992yc}-\cite{Ambjorn:1992aw}. The strong revival of interest since the last decade starting in 2009 with the discovery of complex colored RTMs \cite{Gurau:2011xp}-\cite{Gurau:2013pca}. In contrast with their formers, colored RTMs admit a $1/N$ expansion controlled by the so-called Gurau degree of the corresponding Feynman graphs, which plays the same role as the genus for RMMs. The Gurau degree is reduced to the genus in dimension $2$; and may be defined as the sum of the genera of jackets, which are ribbon subgraphs of tensor diagrams. Interestingly, the Gurau degree is not a topological invariant but allows to properly construct the leading order graphs in the large $N$-limit, called melons \cite{Gurau:2011xq}, in the same way as the planar graphs for RMM. The melons, in dimension $d>2$ corresponds to particular simplicial decomposition (that we abusively call ‘‘triangulation'') of the $d$-dimensional topological sphere $\mathcal{S}_d$. Moreover, they admit a continuum limit, with entropy exponent corresponding to a branched-polymer phase \cite{Gurau:2013cbh}-\cite{Bonzom:2011zz}. Another important step in the development of RTMs was the discovery of the relation between the existence of an internal index symmetry and the $1/N$ expansion. Indeed, uncolored version of the initial complex colored models was introduced in \cite{Bonzom:2012hw}, where the authors highlight the connection between the existence of a $1/N$ expansion and the global $\U(N)^{\times d}$ invariance of the classical action. This simple observation leads to an extension of the colored formalism, in the same universality class \cite{Bonzom:2016dwy}. This connection between symmetry and power counting has been extended a lot, and some models based on $\mathrm{O}(N)^{\times d}$ invariance have been successfully considered, providing triangulation for non-orientable manifolds \cite{Bonzom:2019kxi}-\cite{Bonzom:2019yik}. Some other group have been considered in last years, see \cite{Benedetti:2019ikb}-\cite{Benedetti:2019sop} for the recent reviews. \\
Despite their recent connections with SYK models, and condensed matter physics \cite{Gurau:2016lzk}-\cite{Delporte:2019tof}, RTMs remains essentially, at this day, a promising road for a viable quantum gravity formalism. RTMs arise at the intersection between many current strategies to quantize gravity. Among their inspirations, one has RMMs \cite{Brezin:1992yc}-\cite{Ambjorn:1992gw}, group field theories (GFTs) \cite{Oriti:2009nd} and loop quantum gravity (LQG) \cite{Rovelli:1997yv}-\cite{Rovelli:1998gg};
A recent consequence of GFTs and TMs was the development of tensorial group field theories (TGFTs) \cite{Carrozza:2013wda}-\cite{Lahoche:2015ola}, which improve standard GFT with the tensorial recipe for the construction of their interactions. This was at the origin of some promising renormalization group investigations in the GFTs phase space \cite{Carrozza:2014rba}-\cite{Carrozza:2017vkz}, revealing possible phase transitions compatible with a current scenario about space-time emergence. \\
However, in themselves RTMs admit a rich phase structure. This has been revealed from some analytic investigations conducted in the hope to go beyond the melonic university class and to discover a new continuum limit, with spectral dimension closer to the one of our four-dimensional space-time. To this end, the next-to-leading order (NLO) of the $1/N$ expansion play an essential role. As for matrix models, the double scaling limit for tensor models is based on the observations that NLO contributions become critical at the same point as the melonic contribution. This suggests that we investigate the large $N$ limit and continuum limit in a correlated manner such that we retain the graphs of arbitrary Gurau degree. In the case of matrix models, which can be achieved by sending $N\to \infty$ and $g\to g_c$, the theory is such that the product $N\vert g-g_c\vert^{(2-\gamma)/2}$ (where $\gamma$ is the so-called entropy exponent) remains constant. The same behavior has been achieved for RTMs, in contrast with matrix models such that the double scaling limit leads to a summable series for dimensions less than six. Moreover, the double scaling procedure can be iterated, a multi-critical scenario providing ultimately at the critical point a continuum limit so far from the branched polymer phase of the melonic limit. This multi-scaling scenario remains an attractive and open perspective for RTM. To this end, nonperturbative renormalization group has been envisaged as a promising and alternative way of investigation of this critical behavior; easier than the heavy mathematical machinery used to formally construct the multi-scaling investigations. \\
Using renormalization group to understand critical properties of such a discrete model has been firstly considered in past for matrix models \cite{Higuchi:1993pu}-\cite{Zinn-Justin:2014wva} see also \cite{Higuchi:1994rv}-\cite{Sfondrini:2010zm} for recent results. From the original idea that double scaling limit may be understood as a special parametrization invariance from the long-distance physics along a relevant direction, in complete analogy with what happens for standard critical phenomena. In the references papers \cite{Higuchi:1993pu}-\cite{Zinn-Justin:2014wva} the authors constructed such a renormalization group using perturbation theory and showed the existence of a non-gaussian fixed point with a relevant direction and a critical exponent in qualitative agreement with the analytic calculations of the double scaling limit. The nonperturbative investigations started in \cite{Eichhorn:2013isa}-\cite{Eichhorn:2014xaa} and using Wetterich-Morris formalism \cite{Wetterich:1991be}-\cite{Wetterich:1992yh}, showed significant improvements concerning the perturbative analysis and providing a tractable formalism to explore discrete gravity, and in particular RTMs. The success of this formalism, for matrix models, is because critical exponent for the relevant direction seems to converge toward the exact (analytic) value provided by double scaling limit when the truncation is enlarged. This observation, however, depends crudely on the specific scheme used to compute the critical exponents. This dependence, as pointed out in the reference paper \cite{Eichhorn:2014xaa} could reflect a pathology of the local truncations used to solve the exact renormalization group equations. Indeed, all the considered versions of the nonperturbative renormalization group used a suitable version of the local potential approximation; which of course completely discard the effects arising from the symmetry breaking due to the regulation. This observation is supported by the fact that the heuristic strategy consisting to keep only tadpole diagrams provides the most spectacular convergence toward the exact result, and such a scheme discard strong disagreements with modified Ward identities. The reliability of the method may be checked only because we have the exact result, thus, the question is: {\it can we be confident with the ability of a purely local approximation in the discovering of new multicritical points for tensor models?} Formally, there is no additional difficulty to pass from random matrices to random tensors. The main difference between RMM and RTM in practice is the proliferation of the interactions, and therefore of the beta-functions with the rank of the truncation. Dealing with this difficulty remains tractable for not so large truncations, and the first investigations, as for matrix models, provided encouraging results, (re)-discovering the critical fixed point corresponding to the double-scaling limit, having a single relevant direction with a critical exponent is in qualitative agreement with the exact analytic value $\theta_{\text{exact}}=d-2$. However, as for matrix models, the quantitative agreement depends on the prescription used to compute the critical exponents or the flow equations, and once again is assumed to be a consequence of local truncations. To be used with confident for discovering multicritical points beyond the double scaling limit, the formalism must allow having a control on the approximations, and this is the more important property if we do not have the support of exact analytic calculations to estimate how much the approximation remains physically relevant, or if the discovered critical points are not an artefact of a bad parametrization of the full phase space. Fortunately, some physical guides are allowing to test the reliability of the results obtained in a given prescription without knowledge of the exact solution. The compatibility with constraints arising from symmetries is one of these guides. For random tensor models, the constraints arise essentially from the symmetry breaking due to the regulator; which modify the Ward identities. Purely local potential approximations strongly violate these identities. This fact has been first pointed out in \cite{Lahoche:2018ggd}-\cite{Lahoche:2018oeo}, but the proposed heuristic recipe, taking into account tadpole diagrams to accommodate with Ward identities cannot be used confidently to investigate larger regions of the phase space than the small vicinity of the Gaussian fixed point containing the double-scaling critical point. \\
A systematic analysis of the influence of Ward identities on the behavior of the renormalization group flow has been started for matrix models in \cite{Lahoche:2019ocf}. The authors observed that an approximation scheme solving simultaneously Ward identities and flow equations strongly improve the value of the critical exponent related to the relevant direction, without additional prescription. This is based on the elementary observation that Ward identities and flow equations play a very symmetric role for discrete gravity models \cite{Lahoche:2018ggd}-\cite{Lahoche:2018oeo}, \cite{Lahoche:2020aeh}-\cite{Lahoche:2018vun} see also \cite{Lahoche:2019orv}-\cite{Lahoche:2019cxt} in the case of sixtic interactions and Ising like model. Indeed, the modified Ward identities arise from the symmetry breaking due to the regulator, but this breaking is itself required to construct the RG flow. This is radically different to the situation for ordinary gauge theories, where the RG flow exists independently, due to the non-trivial propagator of the gauge fields, without relation with the symmetry breaking which may arise by introducing the regulator function \cite{Wetterich:2016ewc}-\cite{Safari:2015dva}. For RMM and RTM however, the propagator is trivial, and the symmetry breaking is required to distinguish between UV and IR degrees of freedom. The modification of Ward identities, therefore, is more than a non-trivial aspect of the theory, it is a consequence of the existence of the RG flow itself. With this respect, a violation of Ward identities has to be considered as a serious problem than for ordinary gauge theory. For gauge theory, Ward identities reflect the gauge symmetry, which is unrelated to the scale hierarchy, but it is the case for RMM and RTM. In a recent result, we analyse the flow equations which dictates how to move though scales and also the Ward identities which dictates how to move through momentum space \cite{Lahoche:2020aeh}, and they have been understood as two complementary of the same thing. In \cite{Lahoche:2019ocf}, we proposed two different ways to deal with Ward identities violations and constructed two approximates solutions compatible with them.
The first one was to enlarge the truncation with momentum dependent interaction, reflecting the symmetry breaking. This procedure, as expected, strongly improve the value of the computed critical exponent for the relevant eigendirection at the critical point. However, the presence of derivative couplings, which have non-vanishing value at the fixed point seems to introduce a spurious dependence on the regulator. Such a dependence is, in fact, inevitable in any approximation schemes, and we expect that the sensibility for small deformation of the regulating function may be a good test for the quality of the approximation. From a simple deformation, we showed that the critical exponent does not change significantly around the Litim regulator; in agreement with the familiar claim about its efficiency.
The second strategy was to consider a modified regulator, including fine-tuned counter-terms. These counter terms do not change the UV and IR boundary conditions; and are chosen to cancel the momentum dependent terms in Ward identities using local potential approximation, such that the violation remains as small as possible in the considered range of couplings investigated by the RG flow (expecting that we remain not so far from the Gaussian fixed point, which is essentially the same assumption ensuring the validity of the truncation method). With this method, we found a fixed point, and a critical exponent in very strong agreement with the exact value, ensuring that, up to this fine adjustment of the regulator, the local approximation may be used in practice to solve both flow equations and Ward identities. \\
In this paper, we continue the same analysis for tensor models. We start with complex and real RTMs, having $\U(N)$ invariance, and in both cases, we construct two kinds of approximations, compatible with Ward identities, and investigate the continuum limit through the properties of the resulting fixed points. In detail, the outline is the following:\\
In section \ref{section1}, we recall some basics about RTM, nonperturbative RG formalism and Ward identities. We provide useful definitions and present the notations, as well as the elementary notions, as the proper notion of canonical dimension for RTM. In section \ref{section2} we build a local truncation scheme based on a progressive modification of the (Litim) regulator \cite{Litim:2000ci}-\cite{Delamotte:2007pf}, constructed to cancel the derivative couplings arising from Ward identity, at the order fixed by the truncation. We consider generic melonic truncations up to order eight, and higher truncation in the non-branching sector, and show that the result is systematically better than those obtained from local truncations without modified regulator. We show moreover that connected invariant is insufficient to reach the exact value for the critical exponent and only a local truncation involving the product of local invariants and in agreement with Ward identities allows to converge toward the exact result. In the last subsection, we use the recent effective vertex expansion (EVE) \cite{Lahoche:2018oeo},\cite{Lahoche:2020aeh} to obtain the inductive bound $\theta_{\text{op}}=d-1$ toward which converges the critical exponent for a local truncation of arbitrary order, involving only melonic connected pieces. This shows that independently of the regularization scheme and in agreement with the previous observation that ultralocal approximations do not allow to reach the exact value ($\theta_{\text{exact}}=d-2$). In section \ref{section3} we propose an optimization criterion based on the sensibility of the results under small variations of the regulator and show that physical solutions are systematically stable. In section \ref{section4}, we compare our results with another approximation scheme, including derivative couplings in the truncation, and show that, order by order in the derivative expansion. Including these operators allows to close the (local) infinite hierarchical system of our equation provided by the exact RG equation.

\section{RG flow for U(N) RTMs and local truncations} \label{section1}
In this section we provide some basic material for tensor models and nonperturbative RG formalism. Moreover, we introduce some useful definitions and properties that will be used to construct approximate solutions of the RG equation in the next sections.

\subsection{Wetterich-Morris formalism}
In the Wetterich-Morris formalism, the central object is the effective averaged action $\Gamma_k$, which obeys to the first order flow equation \cite{Wetterich:1991be}-\cite{Wetterich:1992yh}:
\begin{equation}
\frac{\partial}{\partial k} \Gamma_k= \Tr \, \left(\frac{1}{\Gamma_k^{(2)}+R_k} \right)\, \frac{\partial}{\partial k} R_k\,. \label{eq1}
\end{equation}
where $\Gamma_k^{(2)}+R_k$ is the inverse of the effective $2$-point function. The formal trace ‘‘$\,\Tr\,$" depends on the nature of fields involved in the equation and, the regulator function $R_k$ is a scale-dependent mass, chosen such that the degrees of freedom with momentum (smaller than $k$) are frozen out, and discarded from the path integral defining the partition function. For the complex tensor models, this $k$-depends partition function is given by:
\begin{equation}
Z_k[J,\bar{J}]=\int dTd\bar{T}\,e^{-S(T,\bar{T})-\bar{T} R_k T+\bar{J} T+ \bar{T} J}\,, \label{partfunct}
\end{equation}
where:
\begin{itemize}
\item $T$, $\bar{T}$ are respectively the complex tensor and its conjugate, which rank is $d$ and size $N$, $T=\{T_{n_1,\cdots,n_d}\}$, $T_{n_1,\cdots,n_d} \in \mathbb{C}$.
\item The regulator $R_k$ is a diagonal $N^d\times N^d$ matrix, i.e. $(R_k)_{\vec{n}\vec{n}\,^{\prime}}=r_k(\vec{n})\delta_{\vec{n}\vec{n}\,^{\prime}}$; with $\vec{n}:=(n_1,\cdots,n_d)$.
\item The shorthand notations $\bar{T} R_k T$ and $\bar{J} T$ are respectively $\bar{T} R_k T:= \sum_{\vec{n}\vec{n}\,^\prime} \bar{T}_{\vec{n}}(R_k)_{\vec{n}\vec{n}\,^\prime} T_{\vec{n}\,^\prime}$ and $\bar{J} T:= \sum_{\vec{n}} \bar{J}_{\vec{n}} T_{\vec{n}}$.
\item The classical action $S(T,\bar{T})$ is a sum of connected invariant with respect to $U(N)^{\times d}$ transformations.
\end{itemize}
Let us set $R_k=0$. In the viewpoint where the degrees of freedom could be integrated out to build the RG flow, a global unitary invariant theory strongly provides some difficulties, particularly on the degrees of freedom of the initial condition. In standard field theory, we start with UV degrees of freedom, i.e. degrees of freedom having high momenta. This is suitable due to the existence of a nontrivial propagator, which provides a different size for quantum fluctuations. For the tensors models with trivial propagator, all the fluctuations have the same size and we have a canonical notion of UV and IR. The UV being described by the classical action $S(T,\bar{T})$ and the IR, when all the fluctuations are integrated out, by the effective action $\Gamma$:
\begin{equation}
\Gamma[M,\bar{M}]+\ln Z_{k=0}[J,\bar{J}] = \bar{J} M+\bar{M} J\,,
\end{equation}
the classical field $M$ being defined as:
\begin{equation}
M=\frac{\partial }{\partial \bar{J}}\ln Z_k[J,\bar{J}]\,.
\end{equation}
Breaking the unitary invariance, the regulator define a preferred order to make the partial integrations of the degrees of freedom, and provide a path to link UV and IR limits. In the same time, breaking the unitary invariance modify the Ward identities see \cite{BenGeloun:2011xu}-\cite{Perez-Sanchez:2016zbh} and references therein. From the global translation invariance of the partition function \eqref{partfunct} and considering an infinitesimal unitary transformation acting on the first index only, we get:
\begin{equation}
T_{n_1,\cdots,n_d}\to T_{n_1,\cdots,n_d}+\sum_m \epsilon_{n_1m}T_{m,\cdots,n_d}\,,
\end{equation}
with $\epsilon=-\epsilon^\dagger$, leading to the functional equation called the Ward identity given by:
\begin{align}
\nonumber \sum_{\vec{n}_\bot, \vec{n}_\bot^{\prime}}\,\bigg\{ (r_k(\vec{n})&-r_k(\vec{n}^{\prime }))\left[\frac{\partial^2 W_k}{\partial J_{\vec{n}}\partial \bar{J}_{\vec{n}\,^{\prime}}}+\bar{M}_{\vec{n}} M_{\vec{n}\,^{\prime}} \right]\\
&\qquad-\bar{J}_{\vec{n}} M_{\vec{n}\,^\prime}+\bar{M}_{\vec{n}}J_{\vec{n}\,^{\prime}} \bigg\}=0\,.\label{WI}
\end{align}
where $\vec{n}_\bot=(n_2,\cdots,n_d)$. In the limit where the regulator goes to zero, the Ward identity is reduced to:
\begin{equation}
\sum_{\vec{n}_\bot, \vec{n}_\bot^{\prime}}\left(\bar{J}_{\vec{n}} M_{\vec{n}\,^\prime}-\bar{M}_{\vec{n}}J_{\vec{n}\,^{\prime}}\right)=0\,.
\end{equation}
The meaning of this equation can be easily checked taking successive derivative with respect to the external sources. For instance, deriving with respect to $\partial^2/\partial J_{\vec{p}}\,\partial \bar{J}_{\vec{p}\,^{\prime}}$, we get:
\begin{equation}
\delta_{n_1p_1^{\prime}} \,\sum_{\vec{n}^\prime_\bot}\Gamma^{(2)}_{\vec{n}\,^{\prime} \vec{p}}=\delta_{p_1n_1^{\prime}} \, \sum_{\vec{n}_\bot}\Gamma^{(2)}_{\vec{p}\,^{\prime} \vec{n}}\,,
\end{equation}
which is solved by $\sum_{\vec{n}^\prime_\bot}\Gamma^{(2)}_{\vec{n}\,^{\prime} \vec{p}} \propto \delta_{p_1n_1^{\prime}}$, where the coefficient being momentum independent. The same behavior remain true for all colors and we must have: $\Gamma^{(2)}_{\vec{n} \vec{p}} = K \delta_{\vec{n} \vec{p}}$ where $K$ isan arbitrary constant. To be more precise let us notify that the Ward identity arises from the $U(N)^{\times d}$ symmetry ensures that the effective vertices inherit of this invariance as well. The $r_k$ dependent term in \eqref{WI} introduces a momentum dependence, providing a non-zero value for the momentum derivative of effective vertex functions. Moreover, this derivative may be expressed, at the leading order, in terms of generalized trace invariant function, and meaning that such a truncation strongly violates the Ward identity. Obviously, let us consider once again the derivative with respect to $\partial^2/\partial J_{\vec{p}}\,\partial \bar{J}_{\vec{p}\,^{\prime}}$. We get, after some straightforward manipulations:
\begin{align*}
\sum_{\vec{n}_\bot, \vec{n}_\bot^\prime} \bigg\{(r_k(\vec{n})&-r_k(\vec{n}^{\prime })) \left[ G^{(2)}_{\vec{n}\vec{q}} \Gamma_{\vec{q}\vec{q}\,^{\prime} \vec{p}\vec{p}\,^{\prime}}^{(4)} G^{(2)}_{\vec{q}\,^{\prime} \vec{n}^{\prime}} \right] \\
&\qquad+\delta_{\vec{n}\vec{p}} \Gamma_{k,\vec{p}\,^\prime\vec{n}\,^\prime}^{(2)}-\delta_{\vec{n}\,^\prime\vec{p}\,^\prime}\Gamma_{k,\vec{n}\vec{p}}^{(2)}\bigg\}=0\,.
\end{align*}
Let us then simplify this expression. First, due to the momentum conservation along internal faces, we must have $\Gamma^{(2)}_{\vec{n} \vec{p}}= \gamma^{(2)}_k(\vec{p}\,)\delta_{\vec{n} \vec{p}}$. In the same way $G^{(2)}_{\vec{n}\vec{p}}= g^{(2)}(\vec{p}\,) \delta_{\vec{n} \vec{p}}$. Moreover, at the leading order in the large $N$ limit, only the melonic diagram have to be retained for computing the effective loop in the right hand side\cite{Lahoche:2018ggd}-\cite{Lahoche:2018oeo},\cite{Lahoche:2020aeh}. Note with this respect that disconnected pieces does not contribute, as pointed out in \cite{Lahoche:2020aeh}. The melonic contribution $\Gamma_{\vec{q}\vec{q}\,^{\prime} \vec{p}\vec{p}\,^{\prime}}^{(4)}$ must have the following structure:
\begin{equation}
\Gamma_{\vec{q}\vec{q}\,^{\prime} \vec{p}\vec{p}\,^{\prime}}^{(4)}=\sum_{i=1}^d\Gamma_{\vec{q}\vec{q}\,^{\prime} \vec{p}\vec{p}\,^{\prime}}^{(4,i)}\,,
\end{equation}
with:
\begin{equation}
\Gamma_{\vec{q}\vec{q}\,^{\prime} \vec{p}\vec{p}\,^{\prime}}^{(4,i)}=2\pi_k^{(2)}(p_i,q_i) \Sym W^{(i)}_{\vec{q}\vec{q}\,^{\prime} \vec{p}\vec{p}\,^{\prime}}\,, \label{decompquartic}
\end{equation}
and $ \Sym W^{(i)}_{\vec{q}\vec{q}\,^{\prime} \vec{p}\vec{p}\,^{\prime}}:= W^{(i)}_{\vec{q}\vec{q}\,^{\prime} \vec{p}\vec{p}\,^{\prime}}+W^{(i)}_{\vec{q}\vec{p}\,^{\prime} \vec{p}\vec{q}\,^{\prime}}$, where:
\begin{equation}
W^{(i)}_{\vec{q}\vec{q}\,^{\prime} \vec{p}\vec{p}\,^{\prime}}=\delta_{q_ip^\prime_i} \delta_{p_i q_i^\prime} \prod_{j\neq i} \delta_{q_j q_j^\prime} \delta_{p_j p_j^\prime}\,.\label{W}
\end{equation}
Taking $\vec{n}_\bot=\vec{n}_\bot^\prime$, we get, for $n_1=n_1^\prime+1$:
\begin{equation}
r_k(\vec{n})-r_k(\vec{n}^{\prime })= \frac{1}{k} \frac{d}{dx_1} \, r_k(\vec{n})\bigg\vert_{n_1=kx_1}+ \mathcal{O}(1/k)\,,
\end{equation}
where, for large $k$, we have introduced the continuous variable $x_1=n_1/k$. Note that $r_k(\vec{n})$ is assumed to be a function of $\vec{n}/k$. Therefore, keeping only the leading order terms in the large $k$ limit, and setting $\vec{p}_\bot=\vec{p}_\bot^{\,\prime}=\vec{0}_\bot$, $p_1=n_1$, $p_1^\prime=n_1^\prime$ and finally $n_1=1$, $n_1^\prime=0$, we obtain:
\begin{equation}
2\pi_k^{(2)}(0,0)\,\mathcal{L}_2=-\frac{d}{dx_1} \gamma^{(2)}_k(\vec{0}\,)\,, \label{Ward1}
\end{equation}
with:
\begin{equation}
\mathcal{L}_p:=\sum_{\vec{n}_\bot} \frac{dr_k}{dx_1} (\vec{n}_\bot) (g^{(2)})^p(\vec{n}_\bot)\,, \label{defLp}
\end{equation}
where we used the notation $f(\vec{n}_\bot)\equiv f(0,n_2,\cdots,n_d)$. The first derivative of the $2$-point function, therefore may be expressed only in terms of trace invariant functions, up to $1/k$ correction. Interestingly we have the formal relation between this equation and the flow equation \eqref{eq1}. As the flow equation dictates how the coupling change when the scale change, the Ward identity dictates how the coupling change in momentum space. The existence of the momentum dependent flow equation dictated by the Ward identity have the same origin as the scale dependent flow equation which is dictated by the Wetterich equation such that the unitary symmetry breaking provided by the regulator. In the next subsection, we will briefly recall what we call local potential and dimension for RTM, and in the next section we will show that the Ward identity are strongly violated for such a local potential, except for fine adjusted regulator, keeping the $r_k$ depending term on the Ward identities as small as possible. \\
To conclude this section, note that we focus on regulators of the form:
\begin{equation}
r_k(\vec{n})=Zf\left(\frac{\sum_i n_i}{k}\right)\,,
\end{equation}
where $Z$ is the wave function renormalization and $f(x)$ is assumed to be derivable and continuous, and satisfy the following criteria:
\begin{enumerate}
\item $f(x)\to 0$ for $x\to \infty$
\item $f(x)\to \infty$ for $x\to 0$\,. \label{enumerate}
\end{enumerate}
Note that in order to make simple analytic calculations, we focus on the Litim's regulator \cite{Litim:2000ci}:
\begin{equation}
f(x)=\left(\frac{d}{x}-1\right)\theta\left(1-\frac{x}{d}\right)\,, \label{litim}
\end{equation}
where $d$ is the rank of the tensor and $\theta$ the Heaviside step function. Particularly, in the following paper, we consider $d=3$.

\subsection{Locality, dimensionality and melonic diagrams}\label{secdim}

\begin{figure}
\includegraphics[scale=0.9]{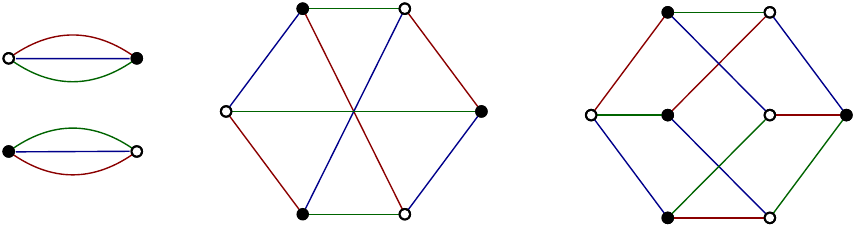}
\caption{Some example of tensorial invariant in rank $3$. The two and third (from left to right) are bubbles. } \label{fig1}
\end{figure}

\begin{figure}
\includegraphics[scale=0.9]{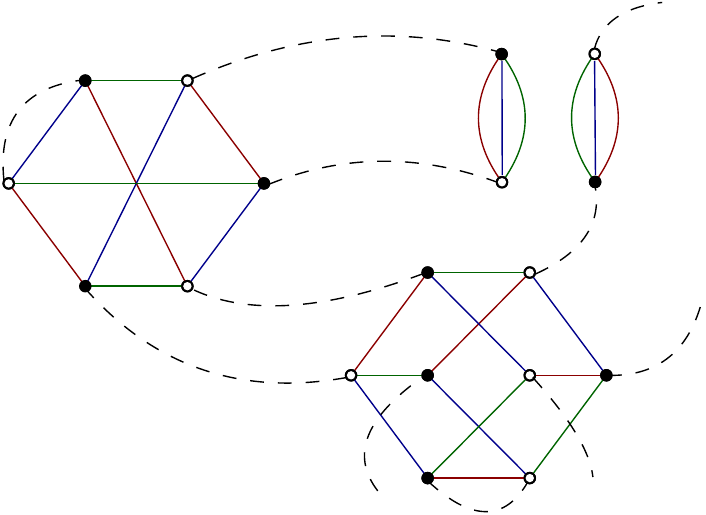}
\caption{A typical Feynman diagram for a three vertex amplitude with four external edges. The dotted edges correspond to Wick contractions.} \label{fig22}
\end{figure}

RTM are non-local theory by construction. Tensorial invariant, whose connected components are called \textit{bubbles} are obtained as the product of the same number of $T$ and $\bar{T}$ fields, contracting their indices pairwise, the index $n_i$ of a field $T$ being contracted with the index $n_i$ of a field $\bar{T}$, ensuring unitary invariance by construction. Usually, these tensorial invariants are pictured as $d$-colored bipartite regular graphs (of rank $d$), where black and white nodes are respectively $T$ and $\bar{T}$ tensors, and the $d$ colored half edges hooked to them represent their $d$ indices. The different following paths which the edges are linked correspond to the invariant contractions. As an example:
\begin{equation}
\vcenter{\hbox{\includegraphics[scale=1]{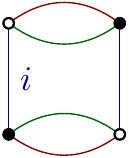} }}=\sum_{\{\vec{p}_j\}}W^{(i)}_{\vec{p}_1\vec{p}_2 \vec{p}_3\vec{p}_4} \, T_{\vec{p}_1}T_{\vec{p}_2}\bar{T}_{\vec{p}_3}\bar{T}_{\vec{p}_4}\,,
\end{equation}
where $W^{(i)}$ has been defined in \eqref{W}. Some examples for rank $3$ are pictured in Figure \ref{fig1}.
The classical action $S(T,\bar{T})$ is assumed to be a local function, admitting an expansion as a sum of bubbles or product of bubbles, weighted by coupling constants $g_{b}$ - labeled by the $d$-colored graph $b$. \\
The connected $2N$-points functions may be expanded in power of couplings, and Feynman amplitudes are indexed by Feynman graphs obtained from Wick theorem. Such a typical diagram is provided on Figure \ref{fig22}. Note that conventionally we associate the color $0$ to the dotted edges. The only change is that Feynman graphs are enriched structure with respect to ordinary graphs, and correspond to the sets of vertices, edges and faces. Before start in detail the calculation of the canonical dimension, let us recall the definition of face:
\begin{definition}
A face is a bi-colored cycle, indexed by a couple $(ij)$. Such a cycle may be open (open face) or closed (closed face).
\end{definition}
The scale behavior is required in the FRG and the scaling means a certain dependence on the number of size $N$ of the component - or equivalently on the dependence on $k$ in the RG flow viewpoint. For RTM, this terminology arises from the existence of a power counting. Indeed, the $1/N$ expansion ensures that, up to a certain rescaling of the coupling constant:
\begin{equation}
g_b=\bar{g}_b N^{\alpha(b)}
\end{equation}
then the Feynman amplitude $A(G)$ for the vacuum Feynman graph $G$ scale as $A(G)\sim N^{d-\frac{2}{(d-1)!}\varpi(G)}$, where $\varpi(G)$ is the Gurau degree given by the following definition:
\begin{definition}
Let $G$ be a $k$-colored bipartite regular graphs with $\vert F \vert$ faces and $p$ black nodes. The Gurau degree $\varpi(G)$ is defined as:
\begin{equation}
\frac{2}{(k-2)!}\varpi(G)=\frac{(k-1)(k-2)}{2} p+(k-1)-\vert F \vert \,.
\end{equation}
\end{definition}
The proper rescaling, for connected tensorial invariants, is given by \cite{Gurau:2011xq}-\cite{Gurau:2013pca}:
\begin{equation}
\alpha(b)=d-1-\frac{2}{(d-2)!}\varpi(b)\,.\label{scaling}
\end{equation}
The leading order diagrams, for which $\varpi(G)=0$ are said to be melonic. For a melonic vacuum diagram, all the interaction bubbles have to be melonic as well. Melonic diagrams may be defined recursively, and their continuum limit corresponds to branched polymer phase \cite{Gurau:2013cbh}. Melonic diagrams with external edges are defined in the same way. They correspond to the leading order diagrams in the $1/N$ expansion, and obey to a similar recursive definition. They can be obtained from vacuum diagrams by deleting some dotted edges.
Locality, in RTM, as in tensor field theories is defined from tensor invariance.
\begin{definition}
The bubbles, or sums of bubbles are said to be ultralocals. Moreover, a sum of bubble and product of bubbles is said to be local.
\end{definition}
From this definition, an effective action builds as sum of bubble is said to be an ultralocal potential. By extension, a local potential involves only bubbles or product of them, including therefore disconnected pieces. The canonical dimension of the interaction arises from the scaling \eqref{scaling}. It is convenient, for RG applications to fix to $1$ the scaling of the kinetic term \cite{Eichhorn:2013isa}-\cite{Eichhorn:2014xaa},\cite{Lahoche:2019ocf}. This can be achieved by a rescaling of the fields: $T\to N^{-\frac{d-1}{2}} T$, modifying the scaling \eqref{scaling} as
\begin{equation}
\alpha^\prime(b)=d-1-\frac{d-1}{2}n(b)-\frac{2}{(d-2)!}\varpi(b)\,.\label{scaling1}
\end{equation}
We call \textit{canonical dimension} this quantity, where $n(b)$ denotes the number of fields involved in the connected bubble $b$. The scaling for bubbles must be completed by the scaling law for disconnected invariants. Let us consider $h=b_1\ast b_2$, a disconnected tensorial invariant made with two bubbles $b_1$ and $b_2$, and we define the difference:
\begin{equation}
\delta \alpha^\prime(b_1\ast b_2)=\alpha^\prime(b_1\ast b_2)-\alpha^\prime(b_1)- \alpha^{\prime}(b_2)\,,
\end{equation}
We fix $\delta\alpha^\prime(h)$ in accordance with the scaling dimension of the kinetic operator, which is set to be zero. To this end, we expect that the scaling of the different operator have to be such that, for any bubble $b$, there exist an optimal way to build a $2$-point diagram $\max G(b)$ whose amplitude $A(\max G)$ scale as $N^0$. The same requirement must be true for interactions made with disconnected pieces. Noting that, with respect to a connected graph, we loss a bicolored cycle merging the two connected components $b_1$ and $b_2$. The resulting graph $G(h)$ can be connected or not. For $G(h)$ being disconnected, we can factorize $ \max G(h)=\max G(b_1) \max \bar{G}(b_2)$, where we assumed that it corresponds to the optimal contraction; and denoted as $\bar{G}(b_2)$ the vacuum graph obtained from $b_2$. From the definition of the amplitude of the graph, we must have
\begin{align*}
A(\max G(h))&\sim A(\max G(b_1))\times A( \bar{G}(b_2))\\
&=\mathcal{O}(1)\times A( \bar{G}(b_2)) \sim N^d\,,
\end{align*}
where we used the scaling theorem for vacuum diagrams. In the case where $\bar{G}(h)$ is connected, the above factorization is not held. However, it is not hard to check that such a contribution have to be less relevant, some internal faces being discarded to ensure connectivity. The optimal counting is therefore $A(\max G(h))\sim N^d$, enforcing to choose (optimally) $\delta \alpha^\prime(h)=-d(k-1)$. In the same way, for a disconnected interaction builds as $k$ bubbles; $h_k=b_1\ast b_2\ast\cdots\ast b_k$, we require $\delta \alpha^\prime(h_k)= -d(k-1)$, and finally:
\begin{equation}
\alpha^\prime(b_1\ast\cdots\ast b_k)=\sum_{\ell=1}^k \alpha^\prime(b_\ell)-d(k-1)\,,
\end{equation}
which can be conveniently rewritten as:
\begin{align}
\nonumber \alpha^\prime(b_1\ast\cdots\ast b_k)&=(d-1)-\frac{d-1}{2}\sum_in(b_i)\\
&\quad -\frac{2}{(d-2)!}\sum_i\varpi(b_i)-(k-1)\,. \label{powercounting}
\end{align}
In the rest of this paper, we denote by $d_g$ the scaling dimension for the coupling $g$. To conclude this part, let us mention that this scaling holds only at zero order around the Gaussian fixed point, as in ordinary quantum field theory; and the first quantum deviations from this Gaussian counting arises from the anomalous dimension. In the present case, it is played by the kinetic prefactor, which we denote by $Z(k)$ (do not make confusion with the partition function $Z_k$) and played the role as an effective mass. It is suitable to set the normalization such that this coefficient remains equal to $1$ along the flow (this is, moreover, a condition to get fixed points). We thus rescale $T$ as $Z^{-1/2}(k) T$, and we finally define the renormalized couplings as:
\begin{equation}
\bar{g}_b= g_b \, Z^{-n(b)/2}(k) N^{-\alpha^{\prime}(b)}\,. \label{rencouplings}
\end{equation}

\subsection{Product of distributions and regularization}
In this manuscript, we will consider the sharp regulators of the form:
\begin{equation}
f(x)=g(x)\theta(1-x)\,. \label{regulatorgeneral}
\end{equation}
on which we intend to give the meaning of the integrals of the form:
\begin{equation}
I_{n,p}=\lim_{\Lambda \to \infty} \int_0^\Lambda \frac{x^n f^\prime(x)}{(1+f(x))^p}\,. \label{typicalintegral}
\end{equation}
Note that this integral appears throughout this paper in the computation of the Wetterich flow equation as well as in the explicit relation of the Ward identities. The
regulator \eqref{regulatorgeneral} introduces a sudden cut in the space of the indices and this is suitable for field theories without background. The Litim regulator \eqref{litim}, commonly used in the FRG literature is an example, with $g(x)=1/x-1$. However, for more general choices of $g(x)$, the exact flow equation \eqref{eq1} cannot be used without a prescription for the product $\delta(x) \theta(x)$; and the integral \eqref{typicalintegral} does not make sense.\\

There are essentially two ways to solve this ambiguity, and we refer respectively to them as ‘‘ scheme 1" ($S_1$) and ‘‘scheme 2" ($S_2$):\\

$\bullet\, S_1:$ In the first scheme, which is the most used in the literature \cite{Morris:1993qb}-\cite{Morris:2000hm}, we solve formally the ambiguity arising in the ill-defined integral \eqref{regulatorgeneral} by considering the distribution of Heaviside as the limit of a regular functions $\theta(x)=\lim_{a\to 0} \theta_a(x)$; for which the integral makes sense. A basic example is:
\begin{equation}
\theta_a(x)=\frac{1}{a\sqrt{\pi}}\int_{-\infty}^x\, e^{-y^2/a^2}dy\,.
\end{equation}
This can also be achieved by a series of functions which converge weakly towards the Heaviside distribution:
\begin{equation}
\Theta_n(x):=\frac{x^n}{e^{x^n}-1}\,,\quad \lim_{n\to \infty} \Theta_n(x)=\theta(x)\,.
\end{equation}
This allows, formally, to remove the ambiguity which appears by rewriting the products like $\delta(x) \theta(x)$. This can be achieved formally from a simple partial integration of \eqref{regulatorgeneral}. Assuming that $f(x)$ is an ordinary regular function rather than a distribution, we have trivially:
\begin{equation}
I_{n,p}= \lim_{\Lambda \to \infty} \left[\frac{n}{p-1} \int_0^\Lambda\, \frac{x^{n-1} dx}{(1+f(x))^p} - \frac{1}{p-1} \Lambda^n \right]\,,\label{typicalintegral2}
\end{equation}
where we assumed that $f(x)$ vanish for large $x$; which is satisfied for a regulator. The two expressions \eqref{typicalintegral} and \eqref{typicalintegral2} are equivalent when $f$ is considered as a function. However, only the last one is well defined when $f$ being a distribution as \eqref{regulatorgeneral}. Then, we can use this form as a definition of the ill-defined product $f'(x) f(x)$ for the computation of the integral. We get explicitly:
\begin{equation}
I_{n,p}=\frac{n}{p-1} \int_0^\alpha\, \frac{x^{n-1} dx}{(1+g(x))^p}- \frac{1}{p-1} \alpha^n\,. \label{reg1}
\end{equation}

$\bullet\, S_2$ In the second scheme, we remember that the derivative $\dot{r}_k$ is a formal operation. Indeed, $k$ must be an integer and $\dot{r}_k$ becomes a formal derivative only in the large $k$ limit. For finite $k$, it must be a finite difference:
\begin{equation}
\dot{r}_k(x) \equiv r_{k+1}(x)-r_k(x)\,, \label{finite}
\end{equation}
and there are no ambiguity with the sums like
\begin{equation}
\mathcal{S}_{n,p}=\sum_{\vec{n}=\vec{0}}^{\infty}\, \big(\sum_i n_i \big)^n \frac{f\left(\frac{\sum_i n_i}{k+1}\right)-f\left(\frac{\sum_i n_i}{k}\right)}{\left(1+f\left(\frac{\sum_i n_i}{k}\right)\right)^p}\,.
\end{equation}
Introducing the parameter $\epsilon:=1/k$, the ambiguity in the formal expression of the product $f'(x) f(x)$ in the integral \eqref{typicalintegral} then writes as:
\begin{equation}
\int x^n\, \frac{\theta(1+\epsilon-x)-\theta(1-x)}{(1+g(x)\theta(1-x))^p} dx\,. \label{finitecont}
\end{equation}
In the interval $x\in [1,1+\epsilon]$, we must have $\theta(1-x)=0$; and in the continuum limit $\epsilon=0$, we may set:
\begin{equation}
\int x^n\, \frac{\theta(1+\epsilon-x)-\theta(1-x)}{(1+g(x)\theta(1-x))^p} dx\to \epsilon \int x^n\, \frac{\delta(1-x)}{(1+‘‘0")^p} dx\,. \label{conv1}
\end{equation}
Note, however that we can make another choice for the finite difference \eqref{finite}. The following example holds
\begin{equation}
\dot{r}_k(x) \equiv r_{k}(x)-r_{k-1}(x)\,, \label{finite2}
\end{equation}
so that the integral \eqref{finitecont} becomes
\begin{equation}
\int x^n\, \frac{\theta(1-x)-\theta(1-\epsilon-x)}{(1+g(x)\theta(1-x))^p} dx\,. \label{finitecont}
\end{equation}
For the ordinary regular functions, there are no difference between left and right derivatives, however in this case, the two definitions are not equivalent at all. In the interval $x\in[1-\epsilon,1]$, we must have $\theta(1-x)=1$, so that with this definition the integral becomes:
\begin{equation}
\int x^n\, \frac{\theta(1-x)-\theta(1-\epsilon-x)}{(1+g(x)\theta(1-x))^p} dx\to \epsilon \int x^n\, \frac{\delta(1-x)}{(1+g(x))^p} dx\,. \label{conv2}
\end{equation}
The convention \eqref{conv2} has been used in the case of matrix models in \cite{Lahoche:2019ocf}. The convention \eqref{conv2} holds for the matrices theories, but becomes pathological for tensors models, with respect to the operations that we will consider for our regulator\footnote{See the next section. Using the first convention, we do not find any solution which makes $\alpha$ such that $\mathcal{L}_2$ vanish.}. Therefore, we keep the second convention given by equation \eqref{conv2} i.e. the scheme 2. \\

Note that except for the case where $g(1)=0$, which corresponds to the Litim regulator; the two definitions are nonequivalents. Moreover let us notify that there are another way to consider the scheme $S_1$ i.e; we can make the restriction on $S_1$ which we will denote by $S_1^\prime$, by starting directly with a regularized expression for the regulator. Indeed, we can use a regularized expression for the Heaviside distribution, $\theta_a(x)$, such that $\lim_{a\to 0} \theta_a(x)=\theta(x)$, to compute the integral \eqref{typicalintegral} \cite{Morris:1993qb}. Indeed, to solve the ambiguity, we have to provide a sense to the limit $\lim_{a\to 0} \theta_a^\prime(1-x) G(\theta_a(1-x))$, for some regular function $G$. This can be achieved for instance using the identity:
\begin{equation}
\theta_a^\prime(1-x) G(\theta_a(1-x))=\frac{d}{dx} \int_{\theta_a(1-x)}^A\,G(y)dy\,,
\end{equation}
where we consider the upper bound $A$, such that the integral exist in the limit $a\to 0$. Taking the limit, it is not hard to show that:
\begin{equation}
\theta_a^\prime(1-x) G(\theta_a(1-x)) \to \delta(1-x)\int_0^1 G(y) dy\,.
\end{equation}
From some elementary algebraic manipulations, it is easy to check that this regularization scheme provides exactly the same expression as \eqref{conv1}; therefore $S_1 \sim S_1^\prime$. \\
In the next sections, we will use these two regularizations schemes, and show explicitly that the corresponding results are strongly dependent on it. This is, once again, an artifact of the symmetry breaking required to construct the RG flow.

\section{Progressive local truncations and modified regulator}\label{section2}

In this section, we construct the approximate solutions of RG equation \eqref{eq1} using local potential approximation. We start with melonic approximation, keeping only connected diagrams. We show that the Ward identity violation can be improved at first order, from an appropriate modification of the Litim regulator without losing its optimal character in the sense of Litim \cite{Litim:2000ci}-\cite{Litim:2001dt}. We then discuss the essential role played by the disconnected diagrams and show that a melonic ultralocal truncation of arbitrary order cannot reach the exact value of the critical exponent for the single relevant direction of the non-Gaussian fixed point. Note that, we keep the notation $d$ for the rank of the tensor, without specifying the value of $d$, to highlight the origin of the contribution, but ultimately we only focus on $d=3$. Moreover, we focus on the \textit{symmetric phase}, and we expand beta functions around vanish means fields $M$ and $\bar{M}$ \cite{Lahoche:2018ggd}-\cite{Lahoche:2019cxt}.

\subsection{Quartic truncation} \label{quartic}

\begin{figure}
\includegraphics[scale=1]{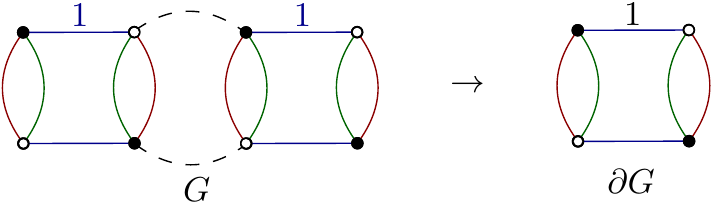}
\caption{On the left a Feynman Graph $G$, and the corresponding boundary graph $\partial G$ on the right.} \label{boundary}
\end{figure}

\begin{figure}
\includegraphics[scale=0.6]{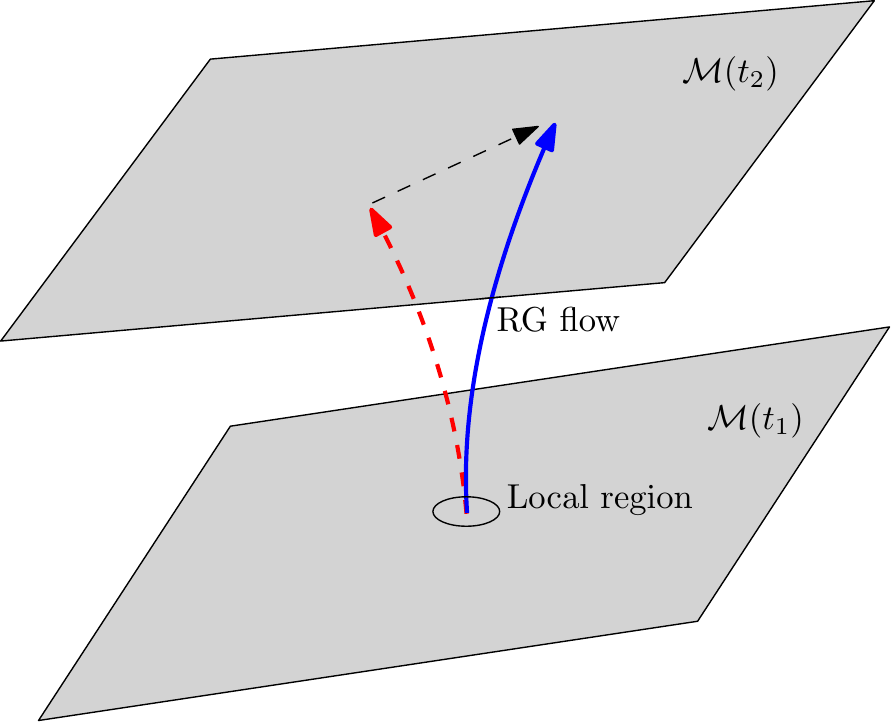}
\caption{Heuristic picture of the RG flow. Starting from a purely local region (corresponding to local truncation), the RG flow (the solid blue arrow) derive toward non local region instead of remain along the red trajectory corresponding to the local flow, due to the Ward identities.} \label{fig3}
\end{figure}

Let us start with a quartic local truncation:
\begin{equation}
\Gamma[M,\bar{M}]= Z(k)\vcenter{\hbox{\includegraphics[scale=1]{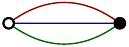} }}+g(k)\,\sum_{i=1}^d \vcenter{\hbox{\includegraphics[scale=0.9]{melon4.pdf} }}\,. \label{truncation4}
\end{equation}
The flow equations can be obtained from the exact flow equation \eqref{eq1} taking successive derivatives with respect to $M$ and $\bar{M}$ fields. The flow equation for $\eta(k)=\partial_k\ln(Z(k))$ can be deduced taking the derivative with respect to $\partial^2/\partial M_{\vec{p}}\, \partial \bar{M}_{\vec{q}}$, and setting $\vec{p}=\vec{q}=\vec{0}$. Graphically, at leading order in $k$ for large $k$, we get an equation of the form:
\begin{equation}
\dot{Z}= -2 g(k) \,\sum_{i=1}^d\vcenter{\hbox{\includegraphics[scale=0.8]{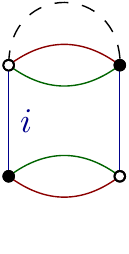} }} \,, \label{dotZ}
\end{equation}
where the dotted edge corresponds to contraction with respect to $\dot{r}_k(G^{(2)})^2$. Moreover, the dot is defined as $\dot{X}= k \partial X/\partial k$ -- the factor $2$ counting the number of derivatives relevant at the leading order in $k$. The flow equation for $g(k)$ may be easily deduced in the same way:
\begin{equation}
d\times \dot{g}=4g^2(k)\sum_{i=1}^d\,\vcenter{\hbox{\includegraphics[scale=0.8]{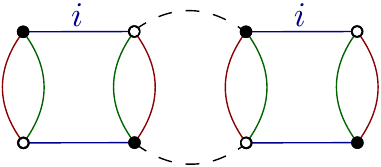} }}\,, \label{dotg}
\end{equation}
where once again the dotted edges represent contractions with propagators $G^{(2)}$ and $\dot{r}_k(G^{(2)})^2$. In principle, we identify the terms on both sides of the flow equations weighting the same boundary graphs. Let us recall the definition of a boundary graph:
\begin{definition}
Let $G$ be a $d+1$-colored Feynman graph (including edges of color $0$) and $\mathcal{F}_{0}$ the set of external faces of type $0i$ for $i\in (1,\cdots,d)$. \\
Let $f_{0i}\in \mathcal{F}_{0}$. The boundary graph of $f_{0i}$ denoted by $\partial f_{0i}$ is the set of bicolored edges of type $0$ and $i$ building the cycle $f_{0i}$. $i$ is called the color of the boundary $/partial f_{0i}$. \\
The boundary graph of $G$, $\partial G$ is a $d$-colored graph (connected or not) build as the set of nodes hooked to external edges and of the boundaries of external faces, such that each boundary $\partial f_{0i}$ is identified to a single edge of color $i$.
\end{definition}
Figure \ref{boundary} provides an illustration of a such boundary graph.
Equations \eqref{dotZ} and \eqref{dotg} can be easily solved using the Litim's regulator. However, the Ward identity \eqref{Ward1} is strongly violated using the Litim's regulator \eqref{litim} with the truncation \eqref{truncation4}. Indeed, the left hand side of equation \eqref{Ward1} writes as
\begin{equation}
-\frac{2}{d} g(k)\sum_{\vec{n}_\bot} \Theta \left(k-\frac{\sum_{i=2}^d n_i}{d}\right) \sim -\frac{2}{d} \,g(k) \, (k\cdot d)^{d-1}\,, \label{Ward1prime}
\end{equation}
where we used the renormalization condition $\pi_k^{(2)}=g(k)$. This term is therefore of order $\bar{g}(k)$; which is in accordance with the expect result. The problem is heuristically pictured on Figure \ref{fig3} where the plane $\mathcal{M}$ represents the largest theory space, including non-local (momentum dependent) couplings. The RG flow thus may be viewed as a map $\mathrm{R}:\mathbb{R}\to \mathcal{M}$, corresponds to the trajectory relying different points of the theory space at different ‘‘times" $t_1=\ln(k_1)$, $t_2=\ln(k_2) \cdots$. Starting with a purely local truncation, involving only bubbles or product of bubbles, the flow does not remain along the local trajectory (the red dotted arrow) but derive toward non-local region. This is a consequence of the Ward identity \eqref{Ward1}. The derivative $d\gamma/dx_1$ being non-zero even if the original truncation involves only the local terms.
To solve this difficulty, and following \cite{Lahoche:2019ocf} we try to modify the windows of allowed momenta, such that:
\begin{equation}
f(x)=\left(\frac{d}{x}-1\right)\theta\left(\alpha-\frac{x}{d}\right)\,, \label{litim3}
\end{equation}
and then fine-tune $\alpha$ such a way that the boundary conditions on $f$ remain holds, and that \eqref{Ward1prime} vanish. Indeed, for $\alpha\neq 1$, the formal derivative of the Heaviside function provides a non-vanishing contribution proportional to $\delta\left(\alpha-\frac{x_\bot}{d}\right)$:
\begin{equation*}
\frac{\partial f}{\partial x_1}\bigg\vert_{x_1=0}=-\frac{d}{x^2_\bot}\theta\left(\alpha-\frac{x_\bot}{d}\right)-\frac{1}{d}\left(\frac{1}{\alpha}-1\right) \delta\left(\alpha-\frac{x_\bot}{d}\right)\,,
\end{equation*}
leading to undefined product as $\delta(\alpha-x)\theta^n(\alpha-x)$. The regularization schemes proposed in the previous section aims to solve this ambiguity. Using the continuum limit, for $k\gg 1$ to replace sums by integrals, we have to compute integrals of the form:
\begin{equation}
J=\int dx_1 dx_2 \theta(\alpha-x_1-x_2) f(x_1+x_2)\,,
\end{equation}
which can be easily computed by elementary algebraic manipulations see Appendix \ref{Appendix}. We get:
\begin{equation}
J=\int_0^\alpha dx\,x f(x)\,.
\end{equation}
Using integral approximation (valid for $d=3$):
\begin{align}
\sum_{\vec{n}_\bot} \left(\frac{\sum_{i=2}^d n_i}{k}\right)^p \theta \left(k-\frac{\sum_{i=2}^d n_i}{d}\right)\approx k^2 \frac{d^{p+2}}{p+2}\,,\label{integralapprox}
\end{align}
and the regularization scheme $S_1$, the condition $\mathcal{L}_p=0$ writes as:
\begin{equation}
\alpha-\frac{1}{p}\alpha^p=0\,,\label{integralp}
\end{equation}
and setting $p=2$, we get $\alpha=2$. Using the scheme $S_2$, and the well known identity $\partial \theta(\alpha-x)/\partial \alpha=\delta(\alpha-x)$, we get straightforwardly an unique solution $\alpha=3/2$\footnote{In rank $d$, it may be easily checked that $\alpha=d/(d-1)$.} $\alpha=3/2$ coming from the equation.
\begin{equation}
-\frac{1}{2}\alpha^2-\alpha^2(1-\alpha)=0\,,
\end{equation}
In the hope to derive other solution for this problem, we can tempting to try the modification $f(x)\to f(x)+\alpha x f(x)$. However, all such a solutions are in conflict with the positivity requirement of the effective propagator in the interval $x\leq d$, which introduces some singularities and therefore, we discarded them.
Finally, let us add an important remark about this derivation. The reader may have some doubts about the use of the truncation \eqref{truncation4} to compute the integral on the left hand side of the Ward identity. However, we have to keep in mind (and it is clear for the regulator that we have chosen) that the windows of momenta relevant for the computation of this integral, provided by the distribution $\partial f/\partial x_1$ is exactly the same ($x_\bot \leq \alpha d$) as the one provided by $\dot{r}_k$ into the flow equation (see equation \eqref{dotr}). Therefore, using the truncation to compute the sum on the right hand side of the Ward identity is not an additional approximation. It is the same as to use the truncation to solve the flow equation.\\
The equations \eqref{dotZ} and \eqref{dotg} can be explicitly computed using the regulator \eqref{litim3}:
\begin{align}
\nonumber \dot{r}_k(kx)=Z\bigg[\eta \left(\frac{d}{x}-1\right)& + \frac{d}{x} \bigg] \theta\left(\alpha-\frac{x}{d}\right)\\
&+Z \left(1-\alpha\right) \delta\left(\alpha-\frac{x}{d}\right)\,, \label{dotr}
\end{align}
where we used $g(x) \delta(\alpha-x)=g(\alpha)\delta(\alpha-x)$. Using the sum \eqref{integralapprox}, we get:
\begin{equation*}
\eta=-6g(k)\frac{k^{d-1}}{Z^2}\left[ \eta\left( \iota_{-1,2}-\iota_{0,2} \right)+\iota_{-1,2}+(1-\alpha)\partial \iota_{0,2}^{(S)}\right]\,,
\end{equation*}
and
\begin{equation*}
\dot{g}=4g^2(k)\frac{k^{d-1}}{Z^2}\left[ \eta\left( \iota_{-1,3}-\iota_{0,3} \right)+\iota_{-1,3}+(1-\alpha)\partial \iota_{0,3}^{(S)}\right]\,,
\end{equation*}
where; in the large $k$ limit:
\begin{equation}
\iota_{p,q}:=\int_0^{\infty} d^{d-1}x \vert x\vert^{q+p} \,\frac{\theta\left(d\alpha-x\right)}{d^{q+p}}=\frac{(\alpha)^{p+q+2}d^2}{p+q+2}\,.
\end{equation}
The explicit expression for $\partial \iota_{p,q}^{(S)}$ however depends on the regularization scheme $S=S_1$ or $S=S_2$. For $S=S_1$ we have:
\begin{equation}
\partial \iota_{p,q}^{(S_2)}=d^2\, \frac{\alpha^2}{p-1}\frac{1}{1-\alpha}
\end{equation}
and for $S=S_2$:
\begin{equation}
\partial \iota_{p,q}^{(S_2)}=\int_0^{\infty} d^{d-1}x \vert x\vert^{q+p} \, \frac{\delta\left(d\alpha-x\right)}{d^{q+p}}=(\alpha)^{p+q+1} d^2\,,
\end{equation}
where the norm $\vert . \vert$ is defined as $\vert x \vert := \sum_i x_i$.
In terms of the renormalized couplings \eqref{rencouplings}, defining $\beta_g:=\dot{\bar{g}}$ the previous equations writes in the scheme $S_1$ as:
\begin{equation}
\beta_g^{(S_1)}=2(1-\eta)\bar{g}+ 12\alpha^4\bar{g}^2 \left[ 3\eta \frac{5-4\alpha}{20}-\frac{3}{4}+\frac{3}{2\alpha^2} \right]\,, \label{flowgS1}
\end{equation}
where
\begin{equation}
\eta^{(S_1)}:=\frac{36 (3-2 \alpha) \alpha^2 \bar{g}}{9 \alpha^3 (3 \alpha-4) \bar{g}-2}\,. \label{etaS1}
\end{equation}
Using the scheme $S_2$:
\begin{equation}
\beta_g^{(S_2)}=2(1-\eta)\bar{g}+ 12\alpha^4\bar{g}^2 \left[ 3\eta \frac{5-4\alpha}{20}+\frac{3}{4}+3(1-\alpha) \right]\,, \label{flowgS2}
\end{equation}
where
\begin{equation}
\eta^{(S_2)}:=-\frac{8}{9 \alpha^3 (3 \alpha-4) \bar{g}-2}-4 \,. \label{etaS2}
\end{equation}
Once again, note that the two schemes are equivalents for $\alpha=1$. Equation $\beta_g=0$ can be exactly solved for arbitrary $\alpha$. In particular:\\

$\bullet$ For $\alpha=1$ (standard Litim regulator); we get two fixed points, $g_1\approx -6.29$ and $g_2\approx -0.037$, with respective anomalous dimension and critical exponents\footnote{We recall that critical exponents are defined as the opposite values of the stability matrix $A_{ij}:=\partial_{g_i} \beta_j$ evaluated at a given fixed point.}:
\begin{equation}
\eta_1\approx -4.14\,;\,\, \theta_1\approx 12.3\,,\quad \eta_2\approx 0.81\,; \,\, \theta_2\approx 2.39\,.
\end{equation}
The first fixed point have a very large critical exponent; and the anomalous dimension violate the \textit{regulator bound}\footnote{The regulator have to be very large in the large $k$ limit. For the Litim regulator, taking into account the definition of $\eta$, we must have $r_k \sim k^{1+\eta}$ in the large $k$ limit; ensuring $\eta>-1$.}. Therefore, at this stage, we do not have confidence with this fixed point, which can be viewed as an artifact of the approximation. \\

$\bullet$ In the scheme $S_2$, for $\alpha=1.5$ we get two fixed points: $g_1^*\approx 1.74$ and $g_2^*\approx 0.017$ with anomalous dimensions and critical exponents respectively:
\begin{equation}
\eta_1^{*}\approx -4.33\,;\,\, \theta_1^{*}\approx 17.0\,,\quad \eta_2^{*}\approx 0.57\,; \,\, \theta_2^{*}\approx 2.26\,.
\end{equation}
The two fixed points have essentially the same characteristics as the fixed point $g_1$ and $g_2$ obtained using the Litim regulator; enforcing the confidence in the local truncation for the existence of this fixed point. The properties of the fixed point $g_2$ coincides with the ones of the relevant fixed point discovered in \cite{Eichhorn:2018ylk} using the same purely local truncation with the Litim's regulator. We see that the modified regulator with $\alpha=3/2$ slightly improves the result; the exact result being $\theta=d-2$ for $d$ is the rank of the tensor \cite{Eichhorn:2018ylk}. \\

\begin{figure}
\includegraphics[scale=0.6]{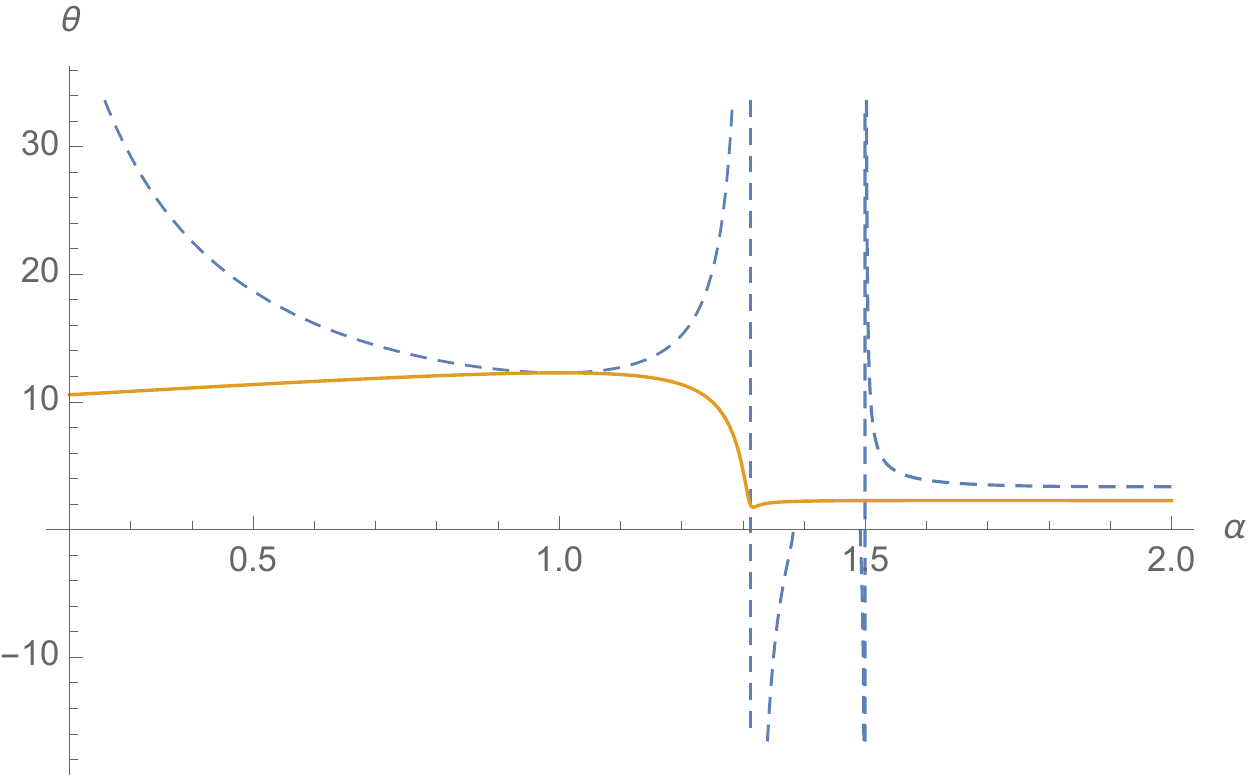}
\caption{Plot of $\theta(\alpha)$ for $\alpha\in[0.2,2]$, using scheme $S_1$ (blue dashed curve) and scheme $S_2$ (yellow solid curve). }\label{thetaalpha}
\end{figure}

$\bullet$ In the scheme $S_1$, for $\alpha=2$, we get again two fixed points, for $g_1^{**} \approx -0.35$ and $g_2^{**} \approx 0.005$, with characteristics:
\begin{equation}
\eta_1^{**}\approx -0.96\,; \,\, \theta_1^{**} \approx 4.96\,;\quad \eta_2^{**}\approx 0.65\,;\,\, \theta_2^{**}\approx 3.35\,.
\end{equation}
The second fixed point $g_2^{**}$ is reminiscent of the two fixed points $g_2$ and $g_2^{*}$; especially concerning the value of their anomalous dimensions, and may be interpreted as the fixed point governing the continuum limit corresponding to the double scaling. However, the first fixed point has the interesting property i.e. the anomalous dimension remains below the lower bound $\eta=-1$. Therefore, there is no reason before discarding it. The only reason may be that: it seems to be very dependent on the scheme used to do the computation; but at this stage, there is no strong indication to privilege scheme $S_1$ regarding the scheme $S_2$. Usually, only the stability regarding higher truncations may provide a solid argument to keep or discard such a fixed point. Nevertheless, stability for small variations of the regulator and the presence of singularities may provide a first indication about the quality of the regularization scheme. Figure \ref{thetaalpha} shows the dependence of the critical exponents for the second fixed point with $\alpha$, respectively for schemes $S_1$ and $S_2$. The blue curve (scheme $S_1$) is stable in the region $\alpha=1$, a possible indication of why the Litim regulator work well\footnote{Note, however that the critical exponents is $\theta\approx 12$ for $\alpha=1$, a characteristic reminiscent of the fixed point $g_1$.}. It becomes stable also in the vicinity of $\alpha=2$, is a larger domain than for $\alpha=1$, better stability which is encouraging physically, despite the strong disagreement with the exact result ($\theta\approx 3.35$ when the exact value is $\theta_{\text{extact}}=1$) -- a conclusion which has to be confirmed for higher truncations. However, between these two regions, the curve of $\theta$ has two singularities. In contrast, the yellow curve in the scheme $S_2$ does not has any singularity. It is stable on a long-range of values around $\alpha=1$; and after a continuous transition, becomes stable once again in the region $\alpha \approx 3/2$. Based on these elementary investigations, scheme $S_2$ seems to behave well than the scheme $S_1$ (in the relevant range of values of $\alpha$ that we investigated); encouraging to take results arising from $S_2$ as reference. We will complete these conclusions in the next sections. \\

\begin{figure}
\includegraphics[scale=0.6]{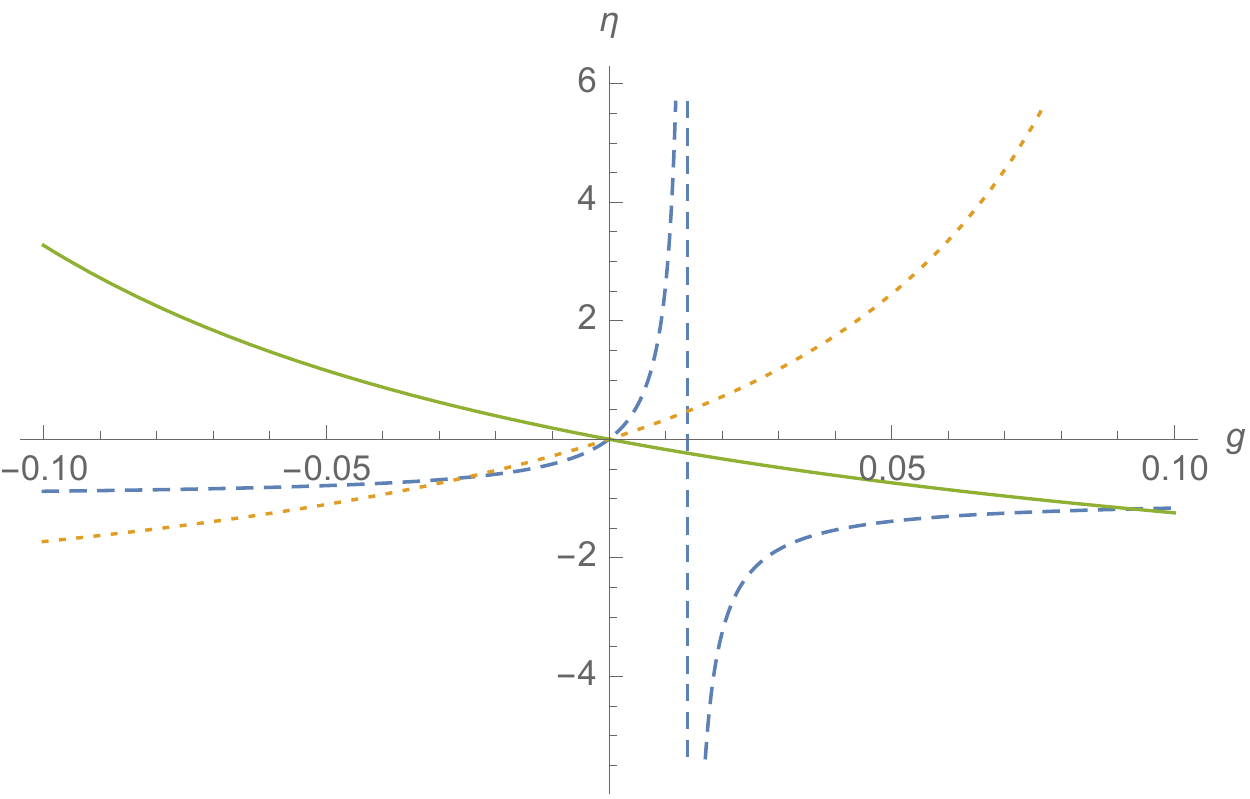} \includegraphics[scale=0.5]{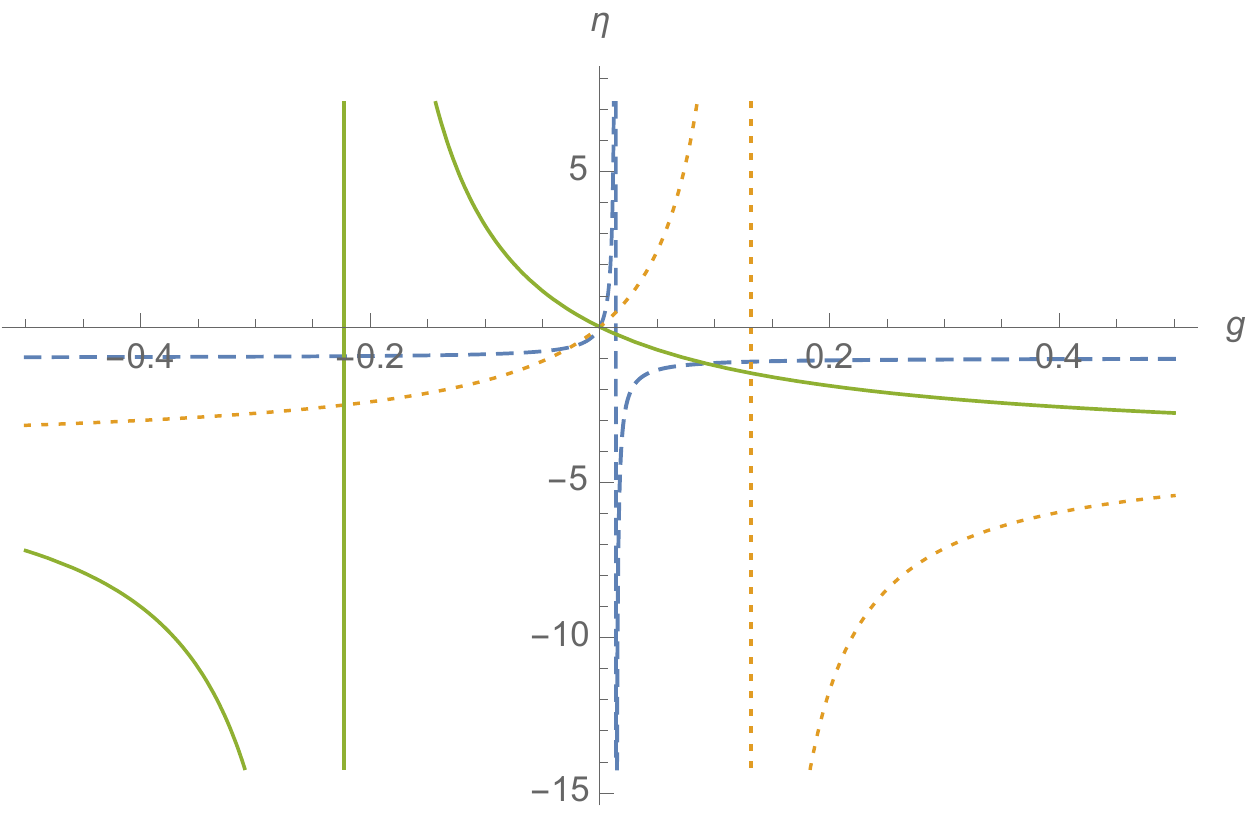}
\caption{Plot of the anomalous dimension in a short range of couplings around the Gaussian fixed point. For $\alpha=1$ (solid green curve), for $\alpha=2$ using scheme $S_1$ (blue dashed curve); and for $\alpha=1.5$ using scheme $S_2$ (dotted yellow curve).}\label{figeta}
\end{figure}

\begin{figure}
\includegraphics[scale=0.6]{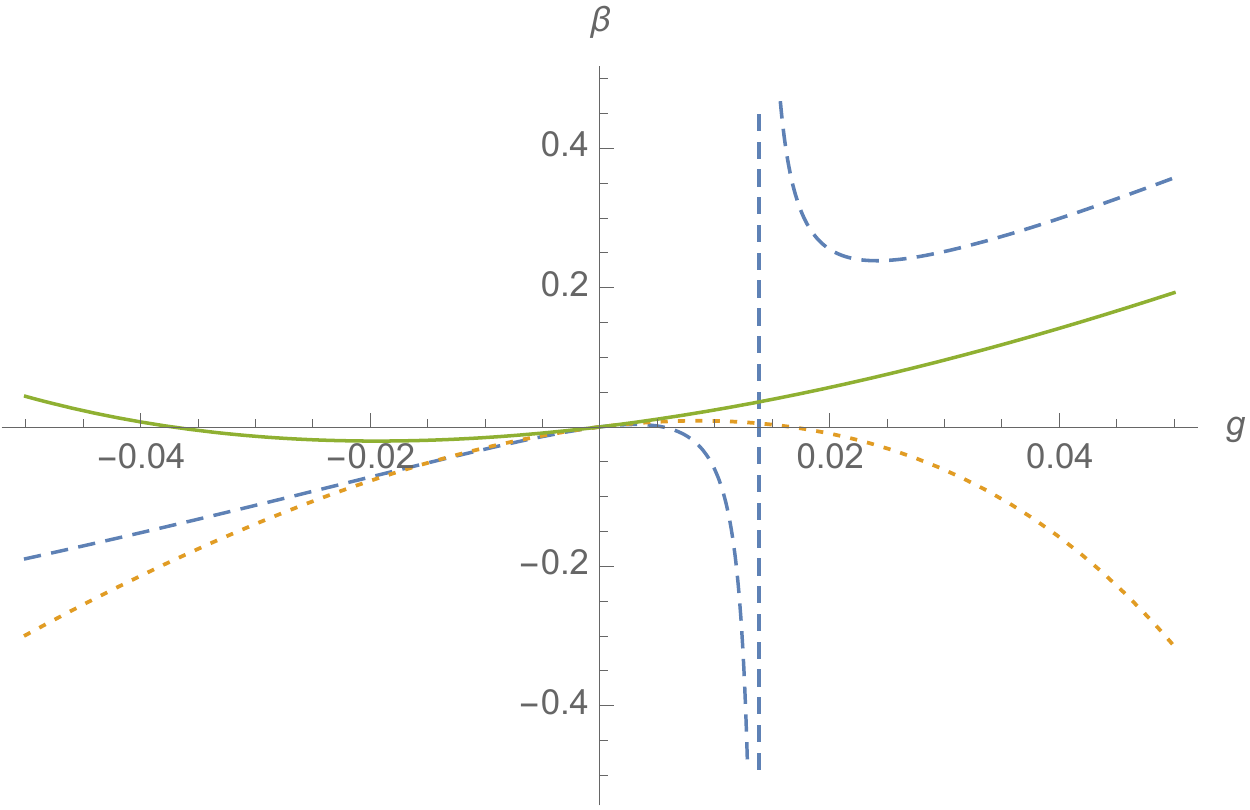}
\caption{Plot of the $\beta$-function for $\alpha=1$ (solid green curve), $\alpha=2$ using scheme $S_1$ (blue dashed curve) and $\alpha=1.5$ using scheme $S_2$ (dotted yellow curve).}\label{beta}
\end{figure}

Figure \ref{figeta} and \ref{beta} show respectively the anomalous dimensions and the $\beta$-functions for $\alpha=1$, $\alpha=2$ and $\alpha=3/2$, respectively using schemes $S_1$ and $S_2$. All the solutions are in quantitative accordance in the vicinity of the Gaussian fixed point, but differ quantitatively and qualitatively in a relatively large range of couplings, before finding a qualitative agreement for couplings of very large magnitude (see the second curve of Figure \ref{figeta} ). Note that all the regularization schemes have a singularity in the vicinity of their zeros; in the negative region for $\alpha=1$ and in the positive region for $\alpha=2$ ($S_1$) and $\alpha=3/2$ ($S_2$). Note that the quality of the regularization scheme could be very dependent on the region that we consider. Indeed, we have seen that, for small couplings, regularization $S_2$ has a better behavior than $S_1$, which is also clear from the curves for $\eta$ and $\beta$, and have a singularity very closer to the Gaussian fixed point. However, the curve for $\eta$ shows that for Litim regulator and $S_2$ approach with $\alpha=3/2$, the anomalous dimension becomes very smaller than the lower bound $\eta=-1$ for couplings with large magnitude. In contrast, the value for anomalous dimension using $S_1$ remains not so far from the bound in the positive region, and just above that in the negative region.

\noindent
It is not easy to say more only from quartic truncations, especially with the improvement coming from taking into account Ward identities in the construction of the local flow. This solution, however, takes into account only the first-order effects, the first derivative for the first Ward identity, involving only $4$ and $2$-point functions. A deeper investigation obviously should take into account higher-order effects. However, we will see in the next section that taking into account first-order effects already shows a clear improvement, mainly visible in the rapidity of the convergence of the results in high truncations.

\subsection{Octic truncations}

\begin{figure}
\includegraphics[scale=0.7]{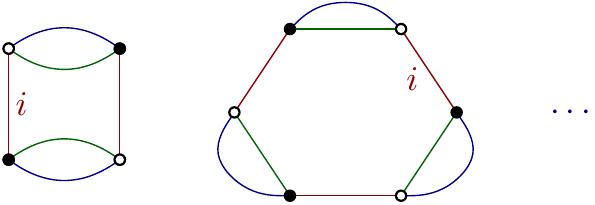} $\,\,\overbrace{\vcenter{\hbox{\includegraphics[scale=0.7]{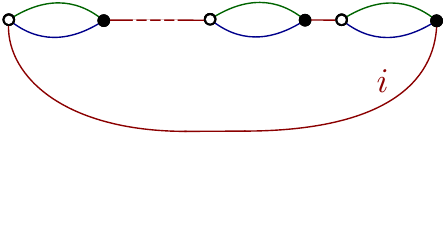} }}}^{p}$
\caption{Structure of the non-branching melon in rank $3$. The last bubble involves $2p$ nodes along the ring of color $i$. We call $2$-dipole the insertions along this mono-colored ring, build as two black and white nodes hooked together by two colored edges of color $\neq i$.}\label{nonbranching}
\end{figure}

In this section we investigate higher order melonic truncations, taking into account sixtic and octic couplings. Taking into account all the melonic connected couplings up to valence eight; we get, in the same notations as in the previous section:
\begin{align}
\nonumber &\Gamma_k[M,\bar{M}]= Z(k)\,\vcenter{\hbox{\includegraphics[scale=1]{melon0.pdf} }}+{g}(k)\,\sum_{i=1}^d \vcenter{\hbox{\includegraphics[scale=0.8]{melon4.pdf} }}\\\nonumber
&+\sum_{i=1}^d \Bigg( {h}_1(k)\vcenter{\hbox{\includegraphics[scale=0.7]{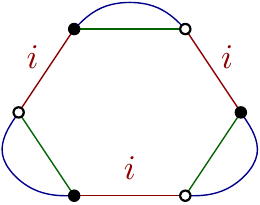} }}+{h}_2(k)\vcenter{\hbox{\includegraphics[scale=0.7]{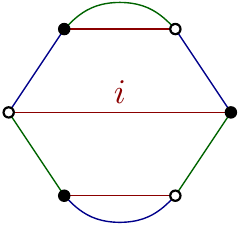} }} \Bigg)\\\nonumber
&+ \sum_{i=1}^d \Bigg({u}_1 \vcenter{\hbox{\includegraphics[scale=0.55]{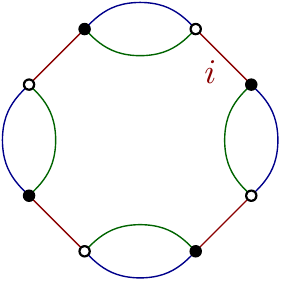}}}+{u}_2 \sum_{j\neq i} \vcenter{\hbox{\includegraphics[scale=0.6]{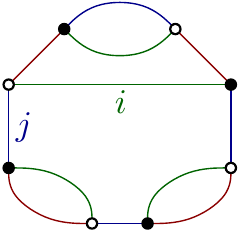}}} \\
&+{u}_3 \sum_{j\neq i} \vcenter{\hbox{\includegraphics[scale=0.6]{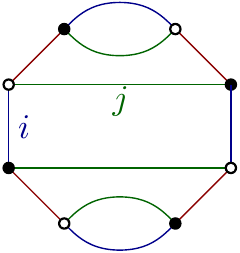}}} +{u}_4\vcenter{\hbox{\includegraphics[scale=0.6]{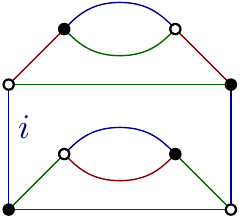}}} +{u}_5 \vcenter{\hbox{\includegraphics[scale=0.6]{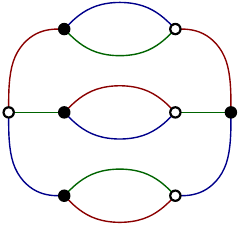}}} \Bigg)\,, \label{truncation8}
\end{align}
The flow equations for the couplings can be easily derived taking successive derivatives of the exact RG equation \eqref{eq1}. The equation \eqref{dotZ} remains unchanged. The equation for $\dot{g}$ however receives sixtic contributions, and becomes graphically
\begin{align}
\nonumber d\times \dot{g}=&4g^2\sum_{i=1}^d\,\vcenter{\hbox{\includegraphics[scale=0.7]{contraction2.pdf} }}-3h_1\sum_{i=1}^d\,\vcenter{\hbox{\includegraphics[scale=0.7]{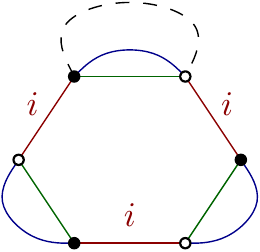} }}\\
&-h_2\sum_{i=1}^d \Bigg( \vcenter{\hbox{\includegraphics[scale=0.7]{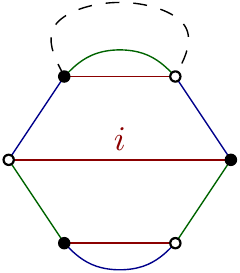} }}+\vcenter{\hbox{\includegraphics[scale=0.7]{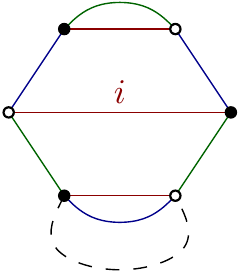} }} \Bigg)\,. \label{dotg2}
\end{align}
The flow equations of ${h}_2$ and ${h}_1$ can be derived in the same way, and we get:
\begin{align}
\nonumber d\times \dot{h}_2= &4gh_2\sum_{i=1}^d\, \left( \vcenter{\hbox{\includegraphics[scale=0.6]{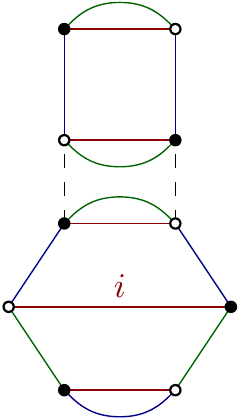} }}+\vcenter{\hbox{\includegraphics[scale=0.6]{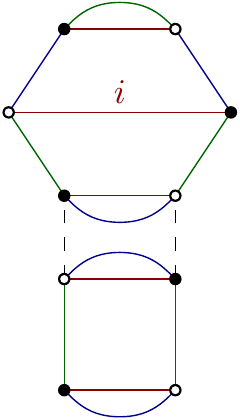} }} \right)\\\nonumber
-&\sum_{i=1}^d \Bigg( u_2\,\sum_{j\neq 1} \, \Bigg(\, \vcenter{\hbox{\includegraphics[scale=0.6]{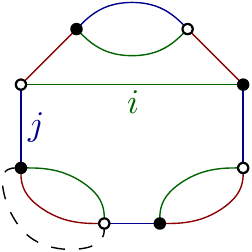}}} +\vcenter{\hbox{\includegraphics[scale=0.6]{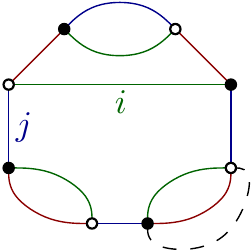}}}\,\Bigg) \\\nonumber
&+u_5\,\vcenter{\hbox{\includegraphics[scale=0.6]{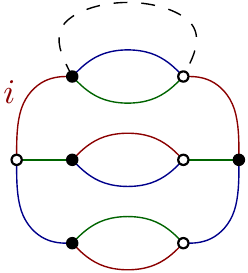}}}
+u_4 \Bigg(\vcenter{\hbox{\includegraphics[scale=0.6]{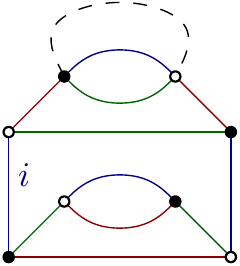}}}+\vcenter{\hbox{\includegraphics[scale=0.6]{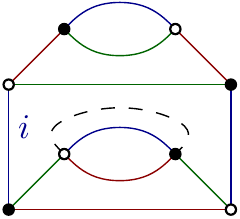}}}\,\Bigg)\\
&+2u_3\,\sum_{j\neq i} \vcenter{\hbox{\includegraphics[scale=0.6]{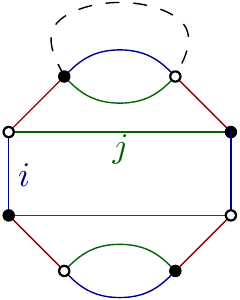}}} \Bigg)\,.\,,
\end{align}
and
\begin{align}
\nonumber d\dot{h}_1&= \, \sum_{i=1}^d\, \left( -8g^3\,\vcenter{\hbox{\includegraphics[scale=0.5]{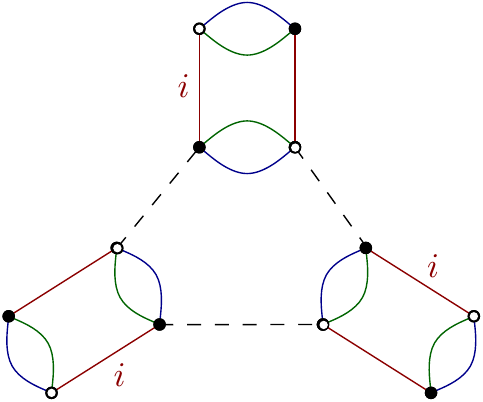} }}+12gh_1\vcenter{\hbox{\includegraphics[scale=0.6]{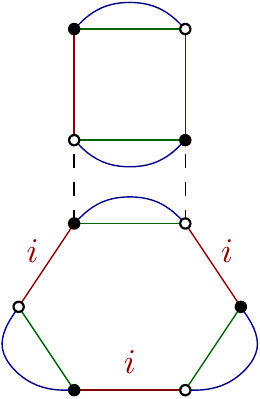}}} \right)\\
&-\sum_{i=1}^d\left(4u_1\,\vcenter{\hbox{\includegraphics[scale=0.6]{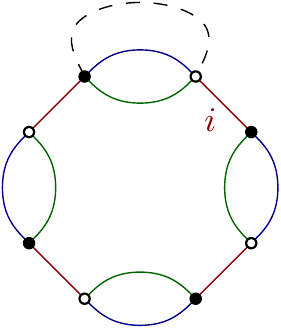}}} +u_2\,\sum_{j\neq 1}\vcenter{\hbox{\includegraphics[scale=0.6]{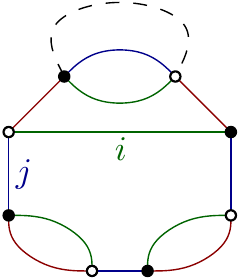}}} \right) \,.
\end{align}
Finally, we get for octic couplings:
\begin{align*}
d \dot{u}_1&= \sum_{i=1}^d \Bigg( 16g^4\, \vcenter{\hbox{\includegraphics[scale=0.5]{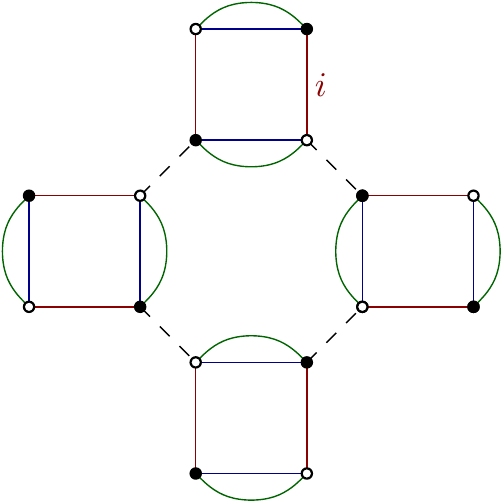}}} -36h_1g^2\,\vcenter{\hbox{\includegraphics[scale=0.5]{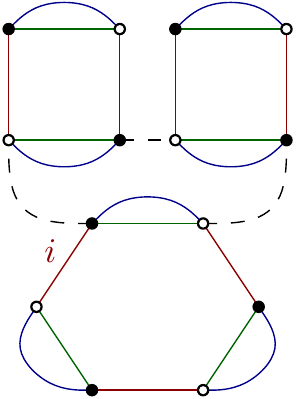}}}\\
&+16 u_1g \,\vcenter{\hbox{\includegraphics[scale=0.5]{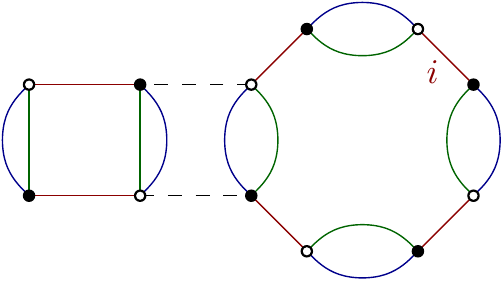}}}\,+9h_1^2\, \vcenter{\hbox{\includegraphics[scale=0.5]{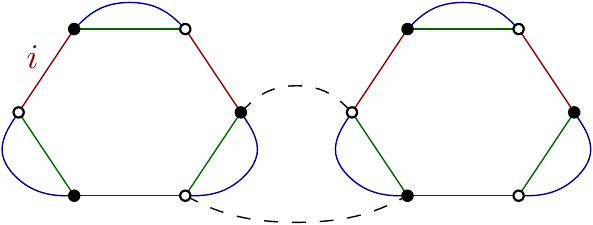}}}\Bigg)\,,
\end{align*}
for $u_1$,
\begin{align*}
&d(d-1) \dot{u}_2= \sum_{i=1}^d\Bigg(-12 h_2g^2\,\sum_{j\neq i} \vcenter{\hbox{\includegraphics[scale=0.5]{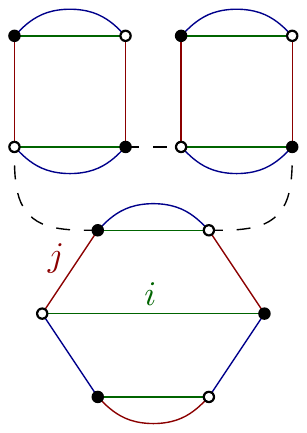}}}\\
&\quad +6h_1h_2\,\sum_{j\neq i}\vcenter{\hbox{\includegraphics[scale=0.5]{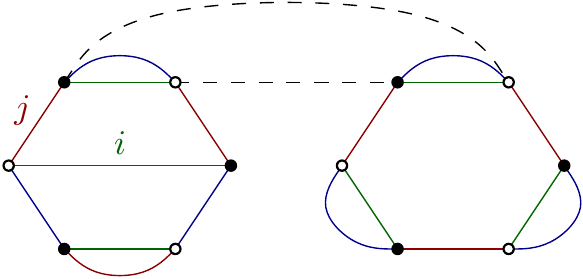}}}+4 gu_2 \sum_{j\neq i} \Bigg( \vcenter{\hbox{\includegraphics[scale=0.5]{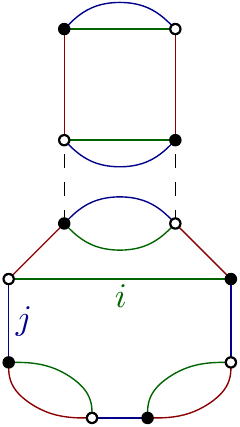}}}\\
&\quad+\vcenter{\hbox{\includegraphics[scale=0.5]{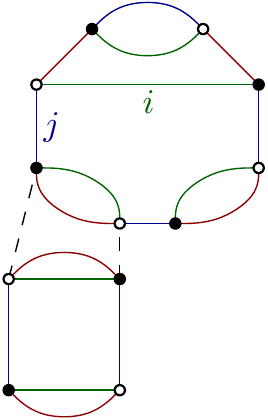}}}+\vcenter{\hbox{\includegraphics[scale=0.5]{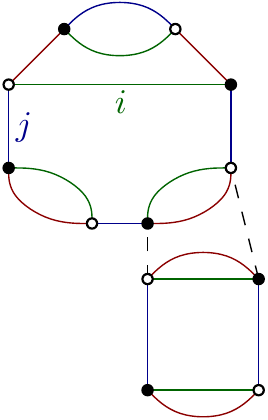}}}\, \Bigg)\Bigg)\,,
\end{align*}
for $u_2$,
\begin{align*}
d(d-1) \dot{u}_3= \sum_{i,j\neq i}\Bigg(8 gu_3 \,&\vcenter{\hbox{\includegraphics[scale=0.5]{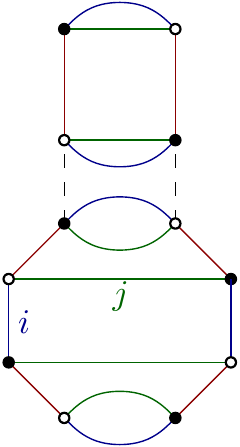}}}\,\\
&+ h_2^2\,\vcenter{\hbox{\includegraphics[scale=0.5]{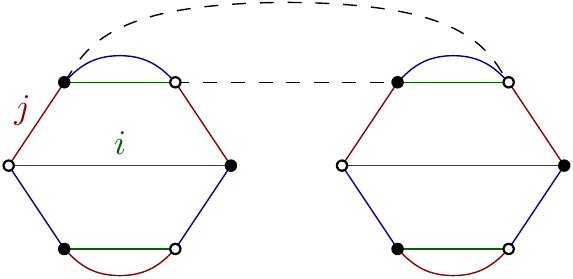}}}\Bigg)\,,
\end{align*}
for $u_3$,
\begin{align*}
d\times \dot{u}_4 = 4 u_4 g \Bigg(\, \vcenter{\hbox{\includegraphics[scale=0.5]{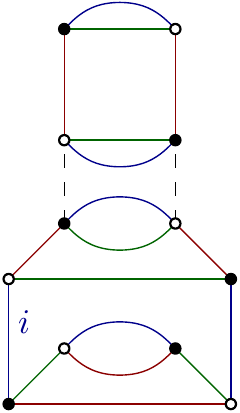}}} +\vcenter{\hbox{\includegraphics[scale=0.5]{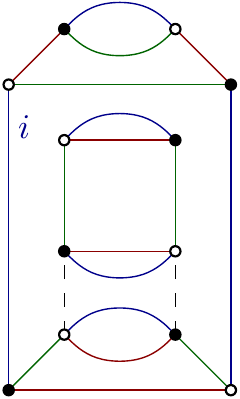}}} \, \Bigg)\,,
\end{align*}
for $u_4$, and finally for $u_5$:
\begin{equation*}
\dot{u}_5= 4 u_5 g \,\sum_{i=1}^d \, \vcenter{\hbox{\includegraphics[scale=0.55]{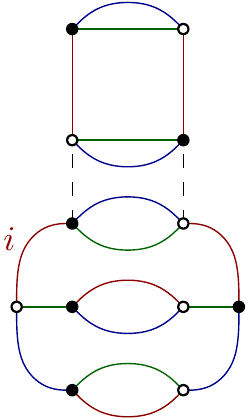}}}\,.
\end{equation*}
These graphical equations may be easily translated in ordinary equations. Defining the scheme dependent symbols $L_n^{(S)}(\eta)$ as:
\begin{equation}
L_n^{(S)}(\eta^{(S)}):= \eta^{(S)}\left( \iota_{-1,n}-\iota_{0,n} \right)+\iota_{-1,n}+(1-\alpha)\partial \iota_{0,n}^{(S)}\,,\label{defL}
\end{equation}
we get straightforwardly:
\begin{align}
\beta_g^{(S)}=&2(1-\eta^{(S)})\bar{g}+4\bar{g}^2(k)L_3^{(S)}-(3\bar{h}_1+2\bar{h}_2) L_2^{(S)}\,,\\\nonumber
\beta_{h_1}^{(S)}=&(4-3\eta^{(S)})\bar{h}_1-8\bar{g}^3L_4^{(S)}+12\bar{g}\bar{h}_1L_3^{(S)}\\\nonumber
&-(4\bar{u}_1+2\bar{u}_2)L_2^{(S)} \,,\\\nonumber
\beta_{h_2}^{(S)}=&(4-3\eta^{(S)})\bar{h}_2+8\bar{g}\bar{h}_2L_3^{(S)}-(2\bar{u}_2+\bar{u}_5+2\bar{u}_4\\\nonumber
& +4\bar{u}_3)L_2^{(S)} \,,\\\nonumber
\beta_{u_1}^{(S)}=&(6-4\eta^{(S)})\bar{u}_1+16\bar{g}^4L_5^{(S)}-36 \bar{h}_1 \bar{g}^2 L_4^{(S)}\\
\nonumber&+(16 \bar{u}_1 g +9\bar{u}_1^2 )L_3^{(S)}\,,\\\nonumber
\beta_{u_2}^{(S)}=&(6-4\eta^{(S)})\bar{u}_2-12 \bar{h}_2 \bar{g}^2L_4^{(S)}+6 \bar{h}_1\bar{h}_2 L_3^{(S)}\\\nonumber
&+12 \bar{g}\bar{u}_2 L_3^{(S)}\,,\\\nonumber
\beta_{u_3}^{(S)}=&(6-4\eta^{(S)})\bar{u}_3+8\bar{g}\bar{u}_3 L_3^{(S)}+\bar{h}_2^2 L_3^{(S)}\,,\\\nonumber
\beta_{u_4}^{(S)}=&(6-4\eta^{(S)})\bar{u}_4+8 \bar{u}_4\bar{g} L_3^{(S)}\,,\\
\beta_{u_5}^{(S)}=&(6-4\eta^{(S)})\bar{u}_5+12 \bar{u}_5 \bar{g} L_3^{(S)}\,;\label{flowgeneral}
\end{align}
the expression for $\eta^{(S)}$ being unchanged:
\begin{equation}
\eta^{(S)}:=-6\bar{g}\,\frac{\iota_{-1,2}+(1-\alpha)\partial \iota_{0,2}^{(S)}}{1+6\bar{g}\left( \iota_{-1,2}-\iota_{0,2} \right)}\,.
\end{equation}
Investigating numerically the fixed points, respectively for sixtic and octic truncations, we get a very large number of solutions. Some of them are irrelevant, violating the regulator bound $\eta=-1$, which seems to be very unstable passing from sixtic to octic truncations. Some of them moreover involve more than one relevant directions, and may be interpreted as multicritical points, corresponding to triple scaling limit and so one \cite{Bonzom:2014oua}-\cite{Gurau:2015tua}. Finally, only one fixed point is physically relevant for double scaling limit, involving only one relevant direction, and have a small dependence on the truncation level. The results for quartic, sixtic and octic interactions are summarized on Figure \ref{table1}. Note that at this fixed point, only couplings $g$, $h_1$ and $u_1$ take a non-zero value. All the other couplings vanish exactly, and the results are essentially insensitive to their presence on the truncation (as we can check explicitly, see the next subsection). This may be viewed as an indication that only a sub-family of melons contribute to the fixed point structure, especially in regard to the understanding of the double scaling limit using renormalization group. This sub-family is known as \textit{non-branching melonic sector}, and non-branching melons may be defined recursively as pictured on Figure \ref{nonbranching}.
We will use this observation in the next section to construct truncations up to order $20$ in the non-branching sector. Another interesting observation can be given by the following prescription: one can mention that the range of values for the couplings at the fixed point seems to follow an interesting hierarchy, $h_1\sim g/10^n$, $u_1\sim h_1/10^n$, $n$ being of order $1$ for standard Litim regulator, and between $1$ and $2$ for regulators with $\alpha=2$ and $\alpha=3/2$, in schemes $S_1$ and $S_2$ respectively. This shows that no significant interacting structure appears up to order $g$.

\begin{figure}[H]
\begin{center}
\begin{tabular}{|c|c|c|c|c|}
\hline
Truncations order &$\alpha$ and scheme & $\bar{g}$ & Relevant $\theta$ & $\eta$\\
\hline
4&$\alpha=1$&-0.04&2.39&0.80\\
\hline
4 & $\alpha=2\,(S_1)$&0.006&3.35&0.65 \\
\hline
4&$\alpha=\frac{3}{2}\,(S_2)$&0.016&2.26&0.57\\
\hline
6&$\alpha=1$&-0.035&2.34&0.71\\
\hline
6 & $\alpha=2\,(S_1)$&0.0046&3.04&0.50 \\
\hline
6&$\alpha=\frac{3}{2}\,(S_2)$&0.01&2.19&0.42\\
\hline
8&$\alpha=1$&-0.03&2.31&0.65\\
\hline
8 & $\alpha=2\,(S_1)$&0.004&2.85&0.40 \\
\hline
8&$\alpha=\frac{3}{2}\,(S_2)$&0.01&2.16&0.35\\
\hline
\end{tabular}
\end{center}
\caption{Characteristics of the non-Gaussian fixed point relevant for the double-scaling limit for octic melonic truncations.}\label{table1}
\end{figure}

From the results summarized on Figure \cite{Bonzom:2014oua}-\cite{Gurau:2015tua}ref{table1}, the following essential observations can be made:
\begin{enumerate}
\item First of all, the characteristics of the fixed point are essentially the same between all the regularization schemes. This concerns both the values of the critical exponents and the values of the anomalous dimension.
\item The values of the relevant characteristics seem to converge toward a finite limit. This is especially the case for the critical exponent, which seems to converge toward $2$. The large truncations that we will consider in the next section for the non-branching sector and the exact results deduced from the EVE method in the next section will confirm this heuristic bound. \\
\item Taking Ward's identities into account may be improved the results qualitatively, compared with the exact results. This is indeed the case for the scheme $S_2$, which effectively improves the result, at the orders considered, compared with the results obtained with the Litim regulator. Besides, the rate of convergence seems significantly faster, for the critical exponent than the anomalous dimension. However, the difference seems worse in the diagram $S_1$, which however displays a speed of convergence greater than the diagram $ S_1 $ when the order of the truncation increases. Besides, the convergence is much slower using the Litim regulator.
\end{enumerate}
From this third point, we deduce that in the point of view of the proximity to the exact result and the speed of convergence with the order of truncation, it seems that any regularization improving the disagreement with Ward's identities will be better than the regular Litim regulator. This result will be supported with large order investigations in the next section. Even to close this part, we aim to add an important remark about the violation of Ward identities. One may object that, even if we fine-tune the regulator to vanish $\mathcal{L}_2$, the disagreement from coefficients that are not canceled could be worse. It is however easy to check that this is not the case. Indeed, from the computations did on the previous section (formula \eqref{integralp}), we seen for instance that using scheme $S_1$, the additional factor $2^p$ arising, setting $\alpha=2$ is compensated by the fact that, numerically $\vert \bar{g}_{\text{Litim}} \vert > 2\bar{g}_{S_1}$, which becomes the tendency that seems to increase with the order of truncation. The same conclusion occurs for the regularization $S_2$. Moreover, the improvement of the hierarchical behavior for higher couplings at the fixed point when $\alpha \neq 1$ enforces this observation.

\subsection{Non-branching sector up to order 20}

\begin{figure}
\includegraphics[scale=0.6]{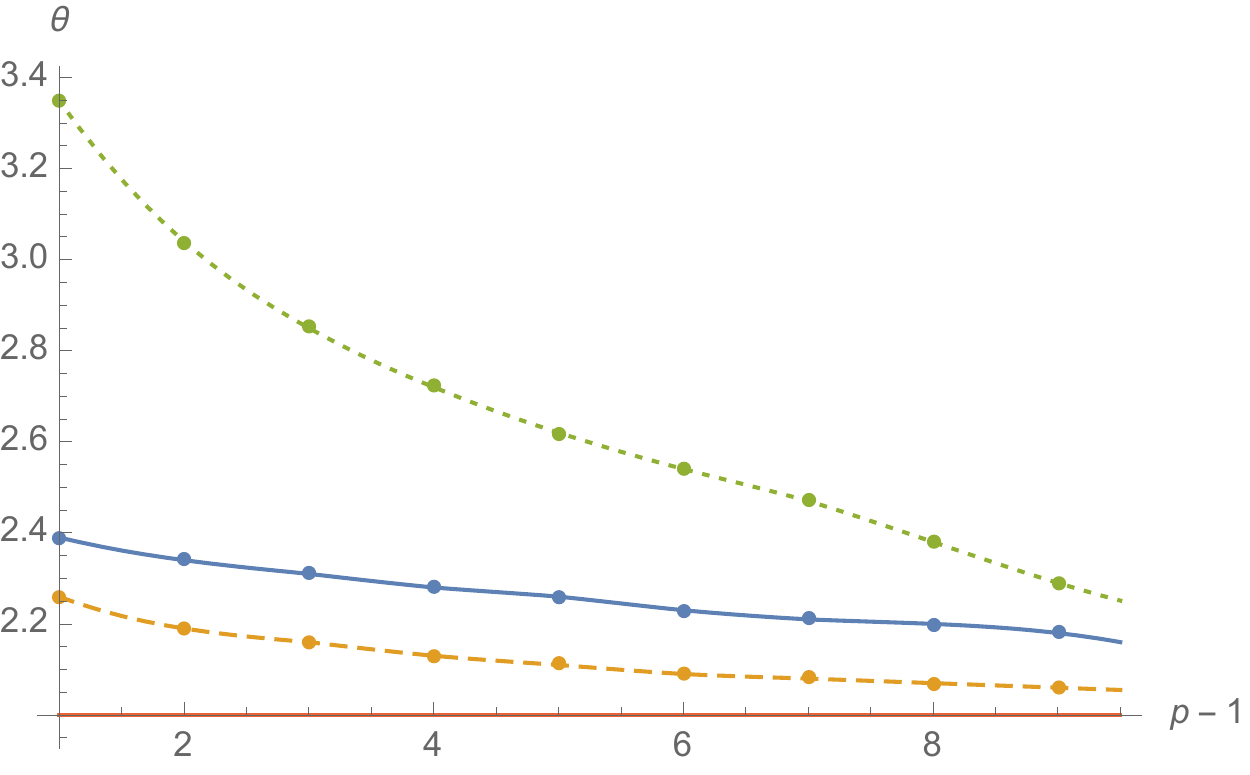}
\caption{The relevant critical exponents in the non-branching sector for truncations up to order $20$. The $x$-axis refers to the order of the truncation $p-1$. The blue (solid) curve is for the standard Litim regulator, the green (dotted) curve is for the scheme $S_1$ ($\alpha=2$) and the yellow (dashed) curve is for the scheme $S_2$ ($\alpha=3/2$).} \label{fig20}
\end{figure}

\begin{figure}
\includegraphics[scale=0.6]{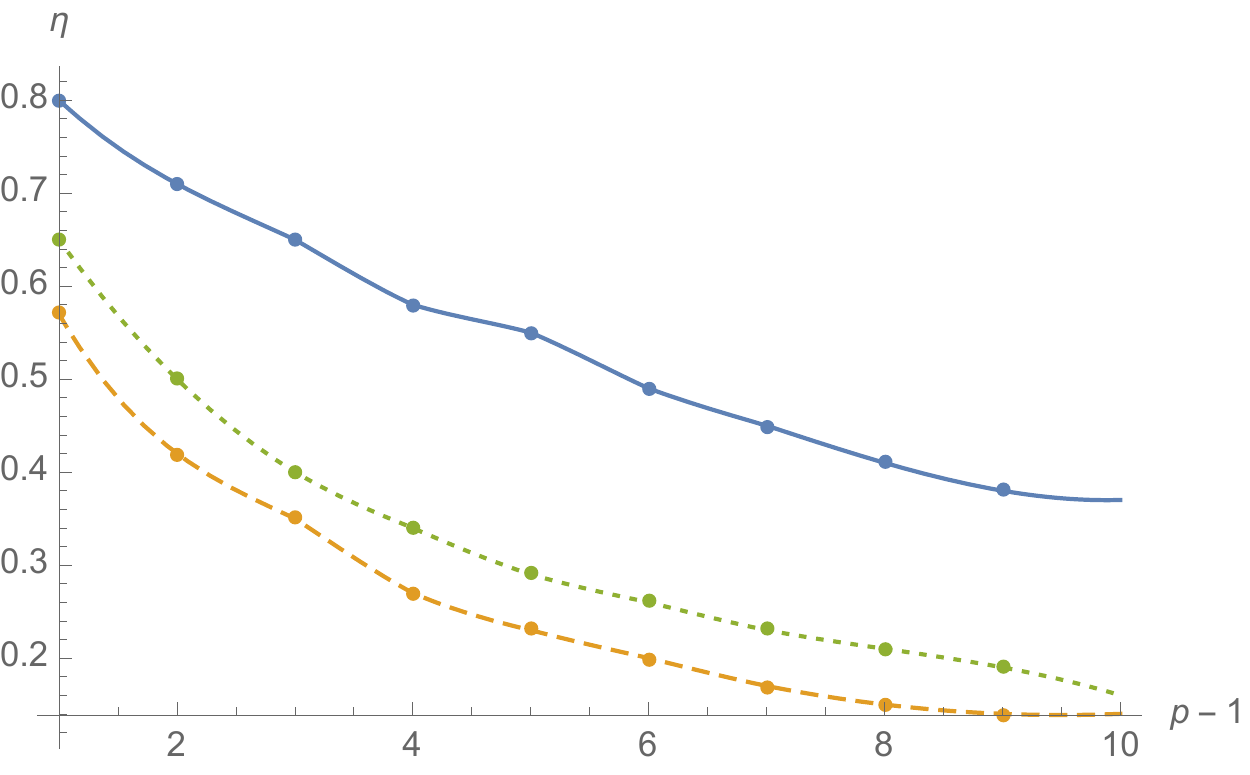}
\caption{The corresponding anomalous in the non-branching sector for truncations up to order $20$. Once again, the $x$-axis refers to the order of the truncation $p-1$. Moreover the color conventions are the same as on the previous Figure: The blue (solid) curve is for the standard Litim regulator, the green (dotted) curve is for the scheme $S_1$ ($\alpha=2$) and the yellow (dashed) curve is for the scheme $S_2$ ($\alpha=3/2$).} \label{fig20bis}
\end{figure}

The non-branching sector obeys to a well know recursive definition, and in this sector, one can easily find an expression for the $\beta$-functions for arbitrary order. For convenience, we introduce the notation $u_{2q}$ for the renormalized coupling (for instance $u_{4}=g$ and so one). It is therefore easy to check recursively that \cite{Carrozza:2016tih}:
\begin{align}
\nonumber \beta_{2p}^{(S)}=&(2(p-1)-p\eta)u_{2p}+\sum_{k=1}^p (-1)^k L_{k+1}^{(S)}(\eta)\\
&\qquad \times \sum_{\{n_{2q}\}\in \mathcal{D}_{k,p}} \frac{k!}{\prod_{q\geq 2} n_{2q}!} \prod_{q\geq 2} (q u_{2q})^{n_{2q}}\,, \label{generalbeta}
\end{align}
where $n_{2q}$ denotes the number of interactions involved on the loop of length $k$; and $\mathcal{D}_{k,p}$ is the set of $\{ n_{2q} \}$ satisfying the two conditions:
\begin{equation}
\sum_{q\geq 2} n_{2q}=k\,,\qquad \sum_{q\geq 2} \,qn_{2q}=p+q\,,
\end{equation}
the first constraint being interpreted as the length of the loop equal to $k$, and the second constraint takes into account that we construct an effective coupling of valence $2p$. Finally, it is easy to count the number of contractions leading to a given non-branching melonic interactions of valence $2p$. Each bubble of type $2q$ involving in a loop has $q$ different positions at the leading order and corresponding to the permutation of the $2$-dipoles along the mono-colored ring. With $n_{2q}$ diagrams, this leads to a factor $q^{n_{2q}}$. Moreover, the $k$ bubbles contributing to the loop of length $k$ can be randomly arranged, for the singular propagator $\dot{r}_k\, (G^{(2)})^2$ (all the other contractions involve only in the effective propagator $G^{(2)}$. The number of arrangement is given by the generalized binomial coefficients:
\begin{equation}
\mathcal{C}_k^{\{n_{2q}\}} \frac{k!}{\prod_{q\geq 2} n_{2q}!}\,,
\end{equation}
and the formula \eqref{generalbeta} follows. We investigated numerically truncations up to order $20$ in this section, and a first observation is that, for the fixed point relevant for double scaling limit, the presence of branching melon has no significant effect on the computation of universal quantities, especially on the values of the critical exponents and anomalous dimension. The results are summarized on Figure \ref{fig20} and \ref{fig20bis}. These figures confirm the assumptions that we have done from octic truncations. On Figure \ref{fig20}, we show that in all cases the value of the critical exponent is improved by the order of the truncation, going more and less rapidly toward the exact value $1$. However, the observed tendency in octic truncation seems to be confirmed. The values progress in direction of the $x$ axis but seems to converge toward $2$ rather than $1$. Despite this disagreement, this value $2$ has a physical meaning. It is nothing but the perturbative result in the first order, and we are tempted to conclude that a purely local truncation cannot significantly improve the physical result, better than the one-loop result. Indeed, the $\beta$--function at one loop must have the following structure:
\begin{equation}
\beta=(d-1)g+ A g^2\,, \label{betaoneloop}
\end{equation}
where $A$ is a constant. This beta function has a fixed point for $g^*=-(d-1)/A$, and the critical exponent writes as:
\begin{equation}
\theta_{\text{one-loop}}= -\beta^\prime(g^*)=-(d-1)-2Ag^*=d-1\,,
\end{equation}
which reduces to $2$ for $d=3$. One expects that for a rank $d$ model, the critical exponent will converge toward $d-1$. We will moreover prove in the section \ref{sectionEVE} that $\theta$ must be equal to $d-1$ for arbitrary large truncations in the non-branching sector using the EVE techniques. Then, remembering that our aim is first to evaluate the quality of the truncation, we have to take the bound $\theta=d-1$ as reference. We will see moreover in the next section that disconnected pieces play an important role if we aim to reach the exact value of $\theta=d-2$. \\

To summarize Figure \ref{fig20}, we conclude that, despite a bad start for the critical exponent using scheme $S_1$ with small truncations,
the speed of convergence is increased when the disagreement with Ward's identities is reduced. The best choice seems to be the scheme $S_2$, which improve both the rapidity of the convergence and the difference to the inductive limit $d-1$ at each order. However, the scheme $S_2$ could compensate this bad, starting by its record convergence speed. The stability of the results for truncations involving a larger family of graphs could be, ultimately, the only way to decide between the reliability of the two schemes. Figure \ref{fig20bis} for the anomalous dimension enforce this conclusion. For the schemes $S_1$ and $S_2$, the anomalous dimension seems to converge rapidly toward a very small inductive limit, when the progression seems to be slowly using Litim regulator.


\subsection{Disconnected pieces}\label{sectiondisco}

In the last sections, we considered connected melonic truncations up to valence $8$ and non-branching sector up to valence $20$. We observed that our results are strongly improved by increasing the order of the truncation, and we expect that this regular progression could converge for sufficiently large truncations. However, we showed that the expected limit does not reach the theoretical result of $\theta=1$, but becomes $\theta=2$. We provide an explanation of this phenomena in the next section. Nevertheless, we completely neglected the influence of next-to-leading order (NLO) bubbles. One may expect that this can be an important mistake for a theory whose interacting fixed point structure arises from the irrelevant couplings. Indeed, from the power counting \eqref{powercounting}, we have seen that non-melonic pieces must have larger canonical dimension than some melonic interactions, and should be included in the truncations. Therefore, for higher-order truncations, one can expect that NLO bubbles could play an important role, especially about the bad melonic limit $\theta=2$. It would not be surprising, moreover, that the NLO contributions play such a role, the double scaling limit being by nature the result which taking into account the influence of the sub-dominant sectors at the critical point. A complete investigation of the influence of sub-dominant orders is reserved for the other article. However, the question of the role of disconnected diagrams is expected to be completely different. Indeed, such a diagram arises for instance from the contraction:
\begin{equation}
\vcenter{\hbox{\includegraphics[scale=1]{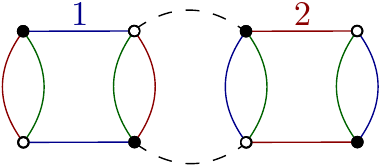} }} \to \vcenter{\hbox{\includegraphics[scale=1]{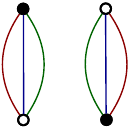} }}
\end{equation}
and does not appears for instance in a truncation involving only a single quartic melonic interaction among the $d$ ones. Therefore, we can in principle discard the influence of disconnected pieces from a breaking of the color symmetry invariance of the models considered above. The results are summarized on Figure \ref{table2}, using scheme $S_2$. We show that the results are significantly closer to the exact limit of $\theta=2$, and, once again, the regulator $\alpha=3/2$ is quantitatively better than the standard Litim regulator. Interestingly, the convergence of the anomalous dimension becomes more precise towards the value $0$; and by using the regulator $\alpha=3/2$ (see claim \ref{claim} of the next section).

\begin{figure}[H]
\begin{center}
\begin{tabular}{|c|c|c|c|c|}
\hline
Truncations order &$\alpha$ (scheme $S_2$) & $\bar{g}$ & Relevant $\theta$ & $\eta$\\
\hline
4&$\alpha=1$&-0.08&2.26&0.57\\
\hline
4&$\alpha=\frac{3}{2}\,(S_2)$&0.036&2.13&0.31\\
\hline
6&$\alpha=1$&-0.06&2.18&0.21\\
\hline
6&$\alpha=\frac{3}{2}\,(S_2)$&0.02&2.08&0.18\\
\hline
8&$\alpha=1$&-0.05&2.15&0.33\\
\hline
8&$\alpha=\frac{3}{2}\,(S_2)$&0.01&2.06&0.14\\
\hline
20&$\alpha=1$&-0.02&2.06&0.13\\
\hline
20&$\alpha=\frac{3}{2}\,(S_2)$&0.005&2.02&0.04\\
\hline
\end{tabular}
\end{center}
\caption{Characteristics of the non-Gaussian fixed point relevant for the double-scaling limit for a single-colored melonic truncation. We considered complete melonic truncations up to order $8$, and only the non-branching sector up to order $20$.}\label{table2}
\end{figure}
As firstly pointed out in \cite{Eichhorn:2017xhy}, the influence of disconnected interactions in not to improve the precision of the critical exponent in regard to the double scaling limit, but to create new fixed points having more than one relevant directions. Such a fixed point is interpreted by the authors as an evidence for a scaling limit beyond double scaling, providing a new continuum limit. However, recovering the double-scaling must seems to require a specific phase space parametrization, breaking the color symmetry. Indeed, we showed that our results in the previous section, taking into account only the melonic sector is in agreement with the simplest truncation breaking the color-symmetry; but, rigorously, we have no reason to discard the disconnected pieces in the color-symmetric truncations, and the relevant fixed point for double scaling disappears. This pathology have been pointed out as a consequence of finite truncations in \cite{Eichhorn:2017xhy}. \\


From the power counting \eqref{powercounting}, an interaction of the form:
\begin{equation}
\left(\vcenter{\hbox{\includegraphics[scale=0.9]{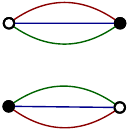} }}\right)\,,
\end{equation}
have canonical dimension $-3$. In contrast, the valence $6$ melonic bubbles have dimension $-2(d-1)=-4$. Therefore, from a strict power-counting point of view, there are no reason to discard disconnected pieces. This also concerns the case for the disconnected piece:
\begin{equation}
\left( \vcenter{\hbox{\includegraphics[scale=0.9]{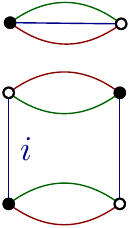} }} \right)\,,
\end{equation}
which has power counting dimension $-5$; smaller than the dimension of melonic octic truncations, which is $-6$.
In this section, we briefly consider their influence.
Let us consider the colored symmetric truncation:
\begin{align}
\nonumber &\Gamma_k[M,\bar{M}]= Z(k)\,\vcenter{\hbox{\includegraphics[scale=0.8]{melon0.pdf} }}+g_1\vcenter{\hbox{\includegraphics[scale=0.8]{disconnected} }}+{g}_2\,\sum_{i=1}^d \vcenter{\hbox{\includegraphics[scale=0.8]{melon4.pdf} }}\\ \nonumber
&+\sum_{i=1}^d \left( {h}_1\,\vcenter{\hbox{\includegraphics[scale=0.7]{int61.pdf} }}+{h}_2\,\vcenter{\hbox{\includegraphics[scale=0.7]{int62.pdf} }} \right)\\
&+\sum_{i=1}^d\left( h_3\, \vcenter{\hbox{\includegraphics[scale=0.7]{disconnectedmelon} }} \right)\,. \label{truncation6bis}
\end{align}
The disconnected terms do not affect the flow equations for $h_1$ and $h_2$ computed in the previous section (setting to zero the octic couplings). The flow equation for $g_2$ receives the additional contribution:
\begin{equation}
-h_3\,\sum_{i=1} \vcenter{\hbox{\includegraphics[scale=0.7]{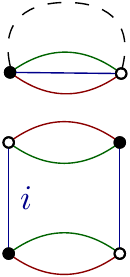} }}\,,
\end{equation}
and the expression of $\dot{Z}$ becomes:
\begin{equation*}
\dot{Z}=-2g_2 \sum_{i=1}^d \vcenter{\hbox{\includegraphics[scale=0.8]{contraction1} }}-2g_1\vcenter{\hbox{\includegraphics[scale=0.9]{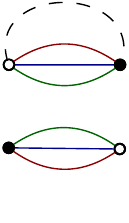} }}\,.
\end{equation*}
Moreover, the couplings $g_1$ and $h_3$ have their own flow equations, explicitly:
\begin{align*}
\dot{g}_1=&4g_1^2\,\vcenter{\hbox{\includegraphics[scale=0.75]{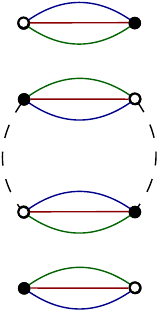} }}+8g_1g_2\sum_i\vcenter{\hbox{\includegraphics[scale=0.7]{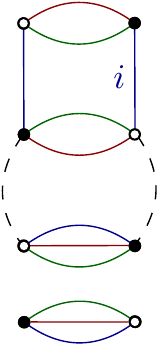} }}+4g_2^2\sum_{i,j\neq i}\vcenter{\hbox{\includegraphics[scale=0.65]{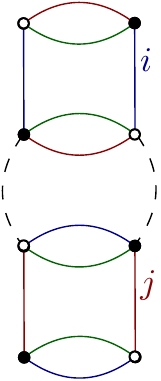} }}\\
&-2h_3\,\sum_{i=1} \vcenter{\hbox{\includegraphics[scale=0.7]{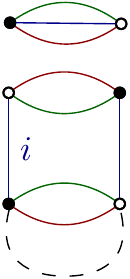} }}
\end{align*}
and
\begin{align*}
&d\dot{h}_3=4g_1 \sum_i \Bigg(h_3\, \vcenter{\hbox{\includegraphics[scale=0.7]{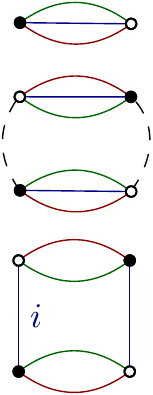} }}+3h_1 \vcenter{\hbox{\includegraphics[scale=0.7]{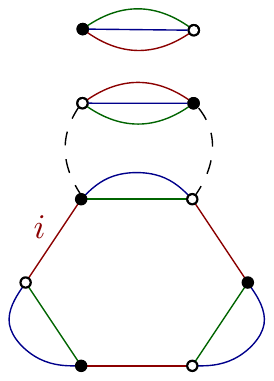} }}\\
&+2h_2 \vcenter{\hbox{\includegraphics[scale=0.7]{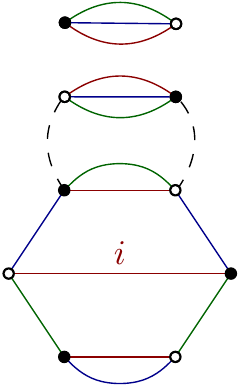} }}\Bigg)+12g_2h_1\sum_{i,j\neq i}\vcenter{\hbox{\includegraphics[scale=0.7]{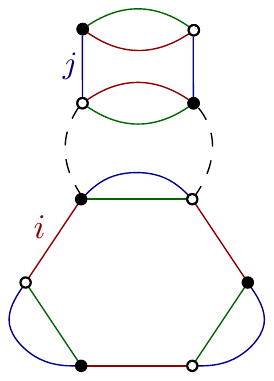} }}\\\nonumber
&+8g_2 h_2\sum_{i,j\neq i}\vcenter{\hbox{\includegraphics[scale=0.7]{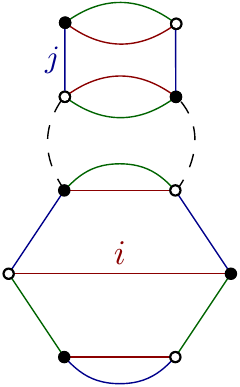} }}+4g_2h_3\sum_{i}\vcenter{\hbox{\includegraphics[scale=0.7]{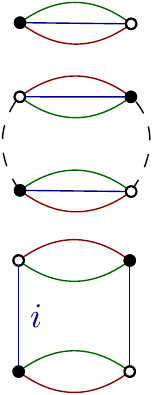} }}\\
&-24\sum_{i=1}^d\,g_2^2 g_1 \vcenter{\hbox{\includegraphics[scale=0.7]{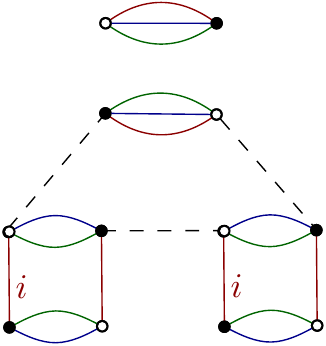} }}-24g_2^3\sum_{i,j\neq i}\vcenter{\hbox{\includegraphics[scale=0.7]{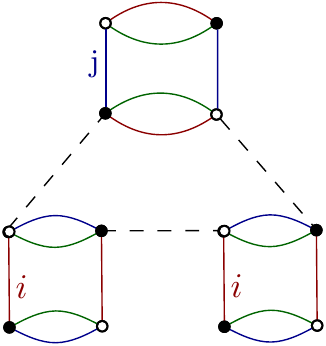} }}\,.
\end{align*}
This leads to the following system:
\begin{align}
\nonumber\beta_{g_1}^{(S)}=&(3-2\eta^{(S)})\bar{g}_1-6\bar{h}_3 L_2^{(S)}+4\bar{g}_1^2K_3^{(S)}+24\bar{g}_1\bar{g}_2 L_3^{(S)}\\\nonumber
&+24\bar{g}_2^2J_3^{(S)}\,,\\\nonumber \beta_{g_2}^{(S)}=&\,2(1-\eta^{(S)})\bar{g}_2+4\bar{g}^2_2L_3^{(S)}-(3\bar{h}_1+2\bar{h}_2) L_2^{(S)}\\\nonumber
&- \bar{h}_3 K_2^{(S)},\\\nonumber
\beta_{h_1}^{(S)}=&(4-3\eta^{(S)})\bar{h}_1-8\bar{g}^3_2L_4^{(S)}+12\bar{g}_2\bar{h}_1L_3^{(S)}\\\nonumber
\beta_{h_2}^{(S)}=&(4-3\eta^{(S)})\bar{h}_2+8\bar{g}\bar{h}_2L_3^{(S)}\,\\\nonumber
\beta_{h_3}^{(S)}=&(5-3\eta^{(S)})\bar{h}_3+4\bar{g}_1\bar{h}_3 K_3^{(S)}+12\bar{g}_1\bar{h}_1 L_3^{(S)}\\\nonumber
&+ 8\bar{g}_1\bar{h}_2 L_3^{(S)}-24\bar{g}_2^2\bar{g}_1L_4^{(S)}-48\bar{g}_2^3 J^{(S)}_4\\
&+16 \bar{g}_2\bar{h}_2J_3^{(S)}+4\bar{g}_2\bar{h}_3K_3^{(S)}\,.
\end{align}
In these equations we introduced $K_n^{(S)}$ defined as:
\begin{equation}
K_n^{(S)}:=\frac{Z^{n-1}}{k^{2}} \int d^dx \frac{\dot{r}_k(x)}{(Z+r_k(x))^n}=:K_n^{(1,S)} \eta^{(S)}+K_n^{(2,S)}\,,
\end{equation}
and the anomalous dimension $\eta^{(S)}$ is given by:
\begin{equation}
\eta^{(S)}=-\frac{6\bar{g}_2 L_2^{(2,S)}+2\bar{g}_1 K_2^{(2,S)}}{1+6\bar{g}_2 L_2^{(1,S)}+2\bar{g}_1 K_2^{(1,S)}}\,,
\end{equation}
where $K_n^{(1)}$ and $K_n^{(2)}$ can be computed exactly as (see Appendix \ref{Appendix}):
\begin{equation}
K_n^{(1)}=\frac{d^3}{2}\left(\frac{1}{n+2}-\frac{\alpha}{n+3}\right)\alpha^{n+2}
\end{equation}
and:
\begin{equation}
K_n^{(2)}=\frac{d^3}{2}\frac{1}{n+2}\alpha^{n+2}+\frac{d^3}{2}(1-\alpha)\alpha^{n+2}\,.
\end{equation}
We moreover introduced the one-dimensional integrals:
\begin{equation}
J_n^{(S)}:=\frac{Z^{n-1}}{k^{2}} \int dx \frac{\dot{r}_k(x)}{(Z+r_k(x))^n}=:J_n^{(1,S)} \eta^{(S)}+J_n^{(2,S)}\,.
\end{equation}
Explicitly:
\begin{equation}
J_n^{(1,S)}=d\left(\frac{1}{n}-\frac{\alpha}{n+1}\right)\alpha^{n}\,,
\end{equation}
and:
\begin{equation}
J_n^{(2,S)}=d\left( \frac{1}{n}+(1-\alpha) \right)\alpha^n\,.
\end{equation}
Note that it is clear that fixed points discarding the disconnected pieces cannot be a fixed point of the previous system. This can be easily checked at the lowest order, keeping only the quartic disconnected pieces (the coupling $g_1$). Setting $g_1=0$, the corresponding flow equation involves the product $d(d-1) g_2^2$, which does not vanish, except for $g_2=0$ or if we consider only one quartic melon among the $d$ allowed (i.e. if we break the color permutation symmetry).

We work only with the scheme $S_2$, and consider the values $\alpha=1$ and $\alpha=3/2$. Starting with the quartic truncation, we get a large number of isolated fixed points. Some of them, however, have to be discarded, violating the regulator bound $\eta=-1$, or being below the singularity line defined by the denominator of $\eta$\footnote{The singularity line defining by the denominator of $\eta$ split the phase space in two connected regions. The denominator is moreover positive only in the region connected to the Gaussian fixed point.}. For $\alpha=1$, the physical relevant fixed point closer to the Gaussian fixed point vanish the branching couplings ($\bar{h}_2=0$), and the critical exponents take the values:
\begin{equation}
\Theta_{\alpha=1}=(5.22, 0.52, -2.40, -1.49)\,,
\end{equation}
with anomalous dimension $\eta_{\alpha=1}\approx 1.14$. For $\alpha=3/2$, we recover a fixed point reminiscent of this one for values:
\begin{equation}
\Theta_{\alpha=3/2}=(8.99, 1.64, -6.84,-2.79)\,,
\end{equation}
and the anomalous dimension $\eta_{\alpha=3/2}\approx 1.44$. Note that in the absence of exact result, we cannot identify which of these results is qualitatively better. The only indication in favor of the second regularization are its good convergence properties in the melonic sector\footnote{Another indication could be the range of values for the couplings at fixed points, sensitively larger for the Litim regulator.}. Moreover, we show explicitly the disappearance of the fixed point with one relevant direction discovered above, illustrating how the results are strongly dependent on the phase space parametrization. \\

Now, let us consider the full sixtic truncation. Once again, we get a large number of isolated fixed points, but only one of them has stables characteristics. We do not recover the fixed points discovered above, but a fixed point having one relevant complex direction. For $\alpha=1$, we get a non-branching fixed point having critical exponent:
\begin{equation}
\Theta_{\alpha=1}=(1.48+0.84i,1.48-0.84i,-4.46, -1.59,-1.30)\,,
\end{equation}
and anomalous dimension $\eta_{\alpha=1}\approx 0.56$. For $\alpha=3/2$ we get:
\begin{equation}
\Theta_{\alpha=3/2}=(2.28+0.68i, 2.28-0.68i,-4.31, -1.95,-1.56)\,,
\end{equation}
with anomalous dimension $\eta_{\alpha=3/2}=0.33$. Once again, we have no reference to compare these results. However, the characteristics of the fixed points obtained from quartic and sixtic truncations seem to be very different. Therefore, a deeper analysis, involving larger truncations is required to conclude about the reliability of this fixed point; or, as for the melonic sector, a deeper understanding of the exact relations between disconnected pieces, as there exist between connected melonic pieces (see the next section).

\subsection{A limit for the ultralocal melonic approximation}\label{sectionEVE}
The previous result showed that the convergence for higher truncation seems to be very dependent on the sectors of the theory space that we take into account. For instance, we showed that taking only the non-branching melonic sector, for instance, cannot allow reaching the exact value of $\theta=d-2$. This result was in a large part of empirical because we only considered three regulators among an infinity of possibilities. In this section we provide a solid argument, based on the effective vertex expansion (EVE), showing that even with a truncation of arbitrarily large size, and without making an explicit choice for the regulator, the critical exponent reaches the value $\theta_{\text{op}}=d-1$, which is nothing but the inductive bound discovered from large truncations in the non-branching sector. \\

Note that in the point of view developed in this paper, the disagreement between the exact value $\theta_{\text{exact}}=d-2$ and $\theta_{\text{op}}$ is not a consequence of the method, but of the restricted domain of the full phase-space that we investigated. We separate the question of exploring vast expanses of phase space from the effectiveness of the method, for which we have retained essentially two criteria, namely proximity with an optimal result (in this case $\theta_{\text{op}}=d-1$) and the speed of convergence. It is expected that the methods giving good results for specific sectors will be as effective on larger domains, and more likely to allow to discover new critical behaviors.\\

EVE is recent development for tensorial group field theories \cite{Lahoche:2018ggd}-\cite{Lahoche:2019cxt}. It allows capturing entire sectors, i.e. an infinite set of effective vertices and their exact momentum dependence, in contrast with crude truncations discussed in the previous section. This method is easy to use only in the non-branching melonic sector, and we only focus on it in this paper. Extensions to sub-leading order is a very fastidious task discussed in \cite{Lahoche:2018oeo}, and no version exist for disconnected interactions. However, the fact that we may able to keep the complete momentum dependence of the effective vertices could strongly improve the local truncation with the Ward-identity violation, without requiring fine-tuning adjustment. We do not discuss in full detail this issue here, referring a more exhaustive analysis to a future article. In section \ref{section4}, we will discuss the influence of derivative coupling, and we will return briefly to the EVE at this time. \\

\begin{figure}
\includegraphics[scale=1]{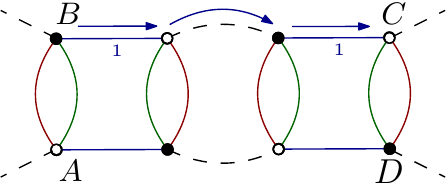}
\caption{A melonic $4$-point graph, with external nodes labeled as $A$, $B$, $C$ and $D$. Pairs $(A,B)$ and $(C,D)$ build $(d-1)$ dipoles. Moreover, pairs $(A,C)$ and $(B,D)$ are boundaries of external cycles of the same colors, one per pair. The corresponding cycle for the pair $(A,C)$ is materialized by the blue arrows.}\label{figprop1}
\end{figure}

\begin{figure}
\includegraphics[scale=0.6]{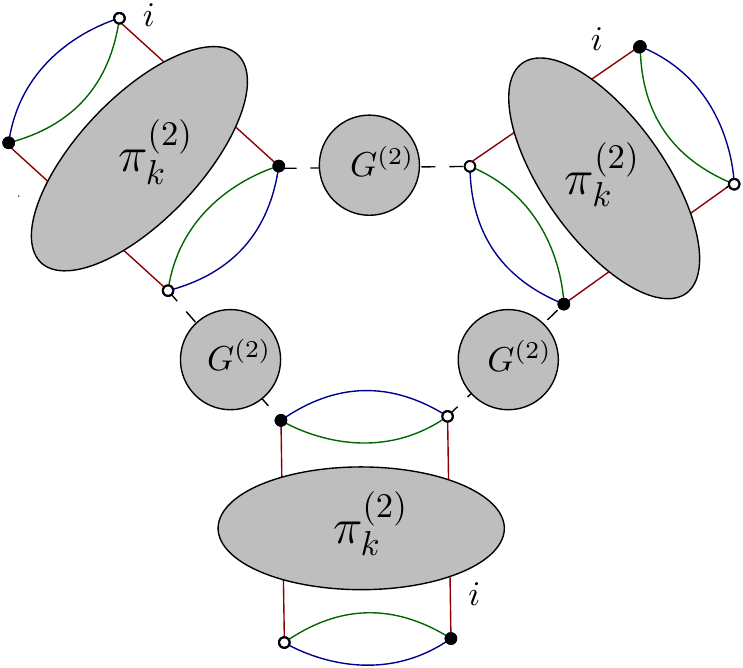}
\caption{Internal structure of the 1PI $6$-points graphs.} \label{figsix}
\end{figure}

Let us consider the quartic model described by the classical action:
\begin{equation}
S(T,\bar{T})= \vcenter{\hbox{\includegraphics[scale=1]{melon0.pdf} }}+ g\, \sum_{i=1}^\nu\vcenter{\hbox{\includegraphics[scale=1]{melon4.pdf} }}\,. \label{classicalEVE}
\end{equation}
Note that we stopped the sum over melonic interaction to the number $1 \leq \nu\leq d$. The reader may be surprised by this restriction. To be more clear we hadn't made an explicit choice of classic action before. We implicitly use the same argument of universality \cite{Delepouve:2014bma}; arguing that the critical behavior of the tensor models must be the same as for the quartic model. We therefore do not lose some thingby restricting ourselves to a quartic model, with which it is easier to work. Investigating the properties of the leading order (i.e. melonics) diagrams, it is not hard to prove the following statement \cite{Lahoche:2018ggd}-\cite{Lahoche:2018oeo}:
\begin{proposition}
Let $G$ be a non-vacuum $1PI$ diagram with $2n$ external edges. The following properties hold:
\begin{itemize}
\item The $2n$ external edges are pairwise connected to $(d-1)$ dipoles. They build $(d-1)n$ open cycles of type $0i$.
\item In addition there exist $n$ open cycles of the same color hooked to external edges pairwise.
\end{itemize}
\end{proposition}\label{prop1}
Figure \eqref{figprop1} provides an illustration of this statement, and we recall the definition of a $k$-dipole:
\begin{definition}
A $k$-dipole is build as two black and white nodes linked together by $k$ colored edges of colors different from $0$.
\end{definition}
As a direct consequence of this proposition, the Feynman graphs involved in the expansion of the effective vertex functions $\Gamma_{k}^{(2n)}$ can be labeled by an index $i$ corresponding to the color of the $n$ open cycles. Thus, $\Gamma_{k}^{(2n)}$ decomposes as as a sum of $d$ functions:
\begin{equation}
\Gamma_{k; \vec{n}_1,\cdots,\vec{n}_2n}^{(2n)}=\sum_{i=1}^d\, \Gamma_{k; \vec{n}_1,\cdots,\vec{n}_2n}^{(2n, i)} \,. \label{melodecomp}
\end{equation}
The Feynman diagrams involved in the expansion of $\Gamma_{k; \vec{n}_1,\cdots,\vec{n}_2n}^{(2n, i)} $ fix completely the relation between the different indices. For $n=2$, the relation between the different indices has been described in \eqref{decompquartic}. Graphically:
\begin{equation}
\Gamma_{k; \vec{q}\vec{q}\,^\prime,\vec{p}\vec{p}\,^\prime}^{(4, i)} = 2\left( \vcenter{\hbox{\includegraphics[scale=0.9]{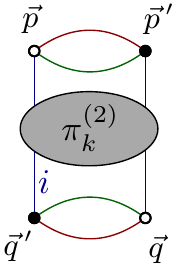} }}+\vcenter{\hbox{\includegraphics[scale=0.9]{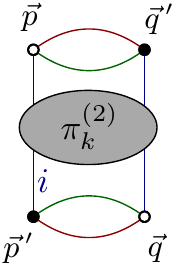} }}\right)\,.
\end{equation}
The aim of the EVE is to close the infinite hierarchical system obtained by expanding the exact flow equation \eqref{eq1}. Restricting firstly our attention to local couplings, this closure requires to express the $6$-point function $\Gamma_{k; \vec{n}_1,\cdots,\vec{n}_6}^{(6, i)}$ in term of the $4$ and $2$-point functions. From proposition \ref{prop1}; the $6$-points vertex function must have the following structure:
\begin{equation}
\Gamma_{k; \vec{n}_1,\cdots,\vec{n}_6}^{(6, i)}=\vcenter{\hbox{\includegraphics[scale=0.7]{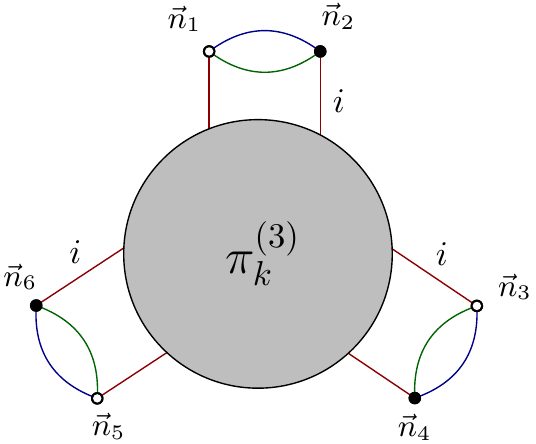} }}+\perm\,,
\end{equation}
where $\perm$ denotes the permutations of the  external momenta. Denoting formally by $\pi_{k}^{(3)}(n_{11},n_{31},n_{51})$ the sum of the interiors of the graphs contributing to the perturbative expansion of $\Gamma_{k; \vec{n}_1,\cdots,\vec{n}_6}^{(6, i)}$. The explicit expressions of $\pi_k^{(2)}$ and $\pi^{(3)}_k$ can be easily obtained from the recursive structure of melonic diagrams \cite{Lahoche:2020aeh}. For $\pi_k^{(2)}$, we get:
\begin{align}
\nonumber \pi_{k}^{(2)}(n,n)&=g(1-2g\mathcal{A}_{2,n}+4g^2(\mathcal{A}_{2,n})^2+\cdots)\\
&=\frac{g}{1+2 g\mathcal{A}_{2,n}}\,, \label{effcoupling}
\end{align}
where:
\begin{equation}
\mathcal{A}_{m,n}:=\sum_{\vec{n}}\, (g^{(2)}(\vec{n}\,))^m \delta_{n_1n}\,. \label{defAp}
\end{equation}
In the same way, the internal structure of the $6$-point melonic diagram can be investigated recursively. The explicit structure is given on Figure \ref{figsix}; which can be translated i as
\begin{equation}
\pi_k^{(3)}(n,n,n)=2(2\pi^{(2)}_k(n,n))^3\,\mathcal{A}_{3,n}\,; \label{sixkernel}
\end{equation}
the combinatorial factor $2$ in front of $\pi^{(2)}_k$ arise from the two allowed orientations for the boundary effective $2$-points vertices. Following \cite{Lahoche:2018vun}, we call \textit{structure equations } the relations \eqref{closed}, \eqref{effcoupling} and \eqref{sixkernel} between effective melonic vertices. Note that, even if we focus on the first relations, such a relation exists to all orders, and the $2n$ point functions may be expressed in term of the $4$ and $2$ point functions\footnote{Note that these structure equations ar nothing but the melonic version of the well-known Schwinger-Dyson equations.} \cite{Lahoche:2020aeh}. Interestingly, all the effective vertices depend on the knowledge of the $2$-point function. This function, or more precisely the self energy $\Sigma_k(\vec{n}\,)$ is determined in the melonic sector from a closed equation. Like the structure equations \eqref{effcoupling} and \eqref{sixkernel}, the closed equation arises directly from proposition \ref{prop1}, and the reader may consult \cite{Lahoche:2015ola},\cite{Samary:2014oya} and references therein.
Defining the mono-colored $2$-point functions $\sigma_k$ as:
\begin{equation}
\Sigma_k(\vec{n}\,):=\sum_{i=1}^d \,\sigma_k(n_i)\,,
\end{equation}
which is nothing but the transcription of equation \eqref{melodecomp} for $2$-point functions, we have the following statement:
\begin{equation}
\sigma_k(n)=-2g \, \sum_{\vec{n}}\, \delta_{n_1n}\, \frac{1}{1-\sum_{i=1}^\nu\sigma_k(n_i)+r_k(\vec{n}\,)}\,. \label{closed}
\end{equation}
We will use this equation especially in section \ref{section4}, investigating the momentum dependence of the melonic functions, in regard to modified Ward identities.
Expanding the flow equation \eqref{eq1} and keeping only the leading order terms in the large-k limit, we get, using the same notations as in the previous section:
\begin{equation}
\dot{\gamma}_k^{(2)}(\vec{n}\,)=-\sum_{i=1}^d\,\vcenter{\hbox{\includegraphics[scale=0.8]{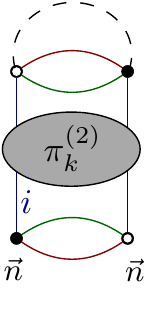} }}\,, \label{eve1}
\end{equation}
and
\begin{equation}
\dot{\Gamma}_{k;\vec{n}\vec{n},\vec{n}\vec{n}}^{(4,i)}= -2\,\vcenter{\hbox{\includegraphics[scale=0.6]{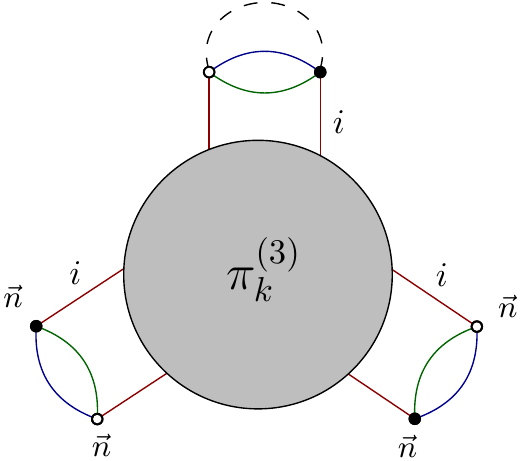} }}+4^2\vcenter{\hbox{\includegraphics[scale=0.6]{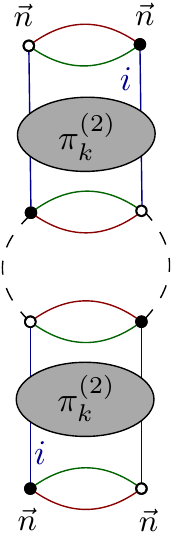} }}\,,
\end{equation}
the factor $2$ in front of the six points contribution arising from the remaining permutation of external edges hooked to white nodes. In this form, the RG equation are completely closed, the $6$-point function being expressed in terms of $2$ and $4$ points ones. Setting the external momenta to zero, and from the renormalization condition
\begin{equation}
\pi_{k}^{(2)}(0,0)\equiv g(k)\,, \label{rencond}
\end{equation}
we deduce:
\begin{equation}
\eta^{(S)}=-2\nu\bar{g}(k) L_2^{(S)}(\eta)\,, \label{etaEVE}
\end{equation}
and:

\begin{align}
\nonumber \beta_{g,EVE}=&((d-1)-2\eta)\bar{g}(k)+4\bar{g}^2(k) L_{3}(\eta)\\
&\qquad -3\, \pi_k^{(3)}(0,0,0) L_{2}(\eta)\,, \label{betaEVE}
\end{align}
where in \eqref{etaEVE} $\nu$ is equal to the number of quartic melonic interactions that we have in the classical action \eqref{classicalEVE}.
These equations are exact, in the sense that they do not required more than exact relations between melonic observables at leading order. At this stage, we can address the following issue: Assuming that we intend to construct an ultra-local approximation of the effective action from an arbitrarily wide truncation, such that this approximation be compatible with the constraints given by the EVE; the $\beta$-function, in an ultra-local truncation has been computed in the previous section, and it corresponds to the previous equation, taking into account the definition $(3!)^2 h_1=(3!)^2 \pi^{(3)}_k(0,0,0)$. However, there are another constraint, arising from the definition \eqref{rencond} of the effective coupling. Indeed, taking the first derivative with respect to $k$, we get, using \eqref{effcoupling}:
\begin{equation}
\dot{g}(k)=4g^2(k) \sum_{\vec{n}}\, \delta_{n_10}\, \frac{- \sum_{i} \frac{d\sigma_k}{dt}(n_i) +\frac{dr_k}{dn} (\vec{n}\,)}{(1-\sum_{i=1}^\nu\sigma(n_i)+r_k(\vec{n}\,))^3}\,.
\end{equation}
Recognizing that $ \frac{d\sigma_k}{dt}$ is nothing but $-\eta/\nu$ in \textit{local approximation}, we thus obtain:
\begin{equation}
\beta_g^{(\text{Exact})}=((d-1)-2\eta) \bar{g} +4\bar{g}^2 L_3(\eta)+ 4\bar{g}^2 \eta \bar{A}_{3,0}\,,\label{betaexact}
\end{equation}
where $\bar{A}_{n,0}$ is the renormalized version of $A_{n,0}$, extracted the global $k$ and $Z$ dependence. Note that these equations only depend on the coupling $g(k)$, which is the only one relevant to drive the RG flow (see \cite{Lahoche:2018oeo},\cite{Lahoche:2020aeh} for more detail). Equation \eqref{betaexact} have to be compared with equation \eqref{betaEVE}. From definition of $\pi^{(3)}_k(0,0,0)$ (equation \eqref{sixkernel}), we conclude that the two equations are compatibles, and the ultra-local melonic approximation makes sense, if:
\begin{equation}
\bar{A}_{3,0}\left( \eta+12\bar{g} L_2 \right)=0\,, \label{compatibilityrel}
\end{equation}
which has two solutions:
\begin{enumerate}
\item $\bar{A}_{3,0}=0$
\item $\eta=-12\bar{g} L_2$\,.
\end{enumerate}
We will investigate separately these two conditions. Note that the second one is in conflict with \eqref{etaEVE} if $\nu\neq 6$. Therefore, if the second condition hold, we have two possibilities: $\nu=6$ or $\nu \neq 6 \Rightarrow \eta=0$. From the second condition, we deduce that equation \eqref{betaEVE} reduces to
\begin{equation}
\beta_g=(d-1)g+4g^2 L_3(\eta=0)\,,
\end{equation}
and, from our perturbative analysis of the previous section (equation \eqref{betaoneloop}) admits a non-Gaussian fixed point for $g^*=-(d-1)/L_3(0)$, with critical exponent:
\begin{equation}
\theta=-\beta_g^{\prime}(g=g^*)=d-1\,.
\end{equation}
The case $\nu=6$ may be analyzed we more attention. Explicitly, the flow equation for the coupling $g$ writes as:
\begin{equation}
\beta_g=((d-1)-2\eta) \bar{g}+4\bar{g}^2 (L_3+\eta \bar{A}_{3,0})\,.
\end{equation}
In a purely local approximation, $L_3(\eta)$ is given by equation \eqref{defL}. To compute $A_{3,0}$, we assume that we work with a sharp regulator ($f$ is proportional to a Heaviside distributions), so that we have:
\begin{equation}
\bar{A}_{n,0}=\sum_{\vec{n}} \delta_{n_10} \left[\frac{\theta\Big(\alpha k-\sum_i n_i\Big)}{(1+f(\vec{n}\,))^n} +\theta\Big(\sum_i n_i- \alpha k\Big)\right]\,.
\end{equation}
The $f$-dependent part of this equation can be easily computed, and we introduce the notation:
\begin{equation}
\mathcal{S}_n:=\sum_{\vec{n}} \delta_{n_10}\frac{\theta\Big(\alpha k-\sum_i n_i\Big)}{(1+f(\vec{n}\,))^n}\,.
\end{equation}
Note that the $f$-independent contribution is $n$-independent as well. Therefore $\bar{A}_{n,0}-\bar{A}_{m,0}=\mathcal{S}_n-\mathcal{S}_m$ is finite. Moreover, from \eqref{effcoupling}, we have:
\begin{equation}
A_{2,0}=\frac{1}{2}\left(\frac{1}{g(k)}-\frac{1}{g(\Lambda)}\right)\,,
\end{equation}
for some $UV$ cut-off $\Lambda$ ($g(\Lambda)$ being what we denoted by $g$ in equation \eqref{effcoupling} for instance, i.e. the initial bare coupling). As a result, we get, after some algebraic manipulations:
\begin{equation}
\bar{A}_{3,0}=\mathcal{S}_{3}-\mathcal{S}_2+\frac{1}{2}\left(\frac{1}{\bar{g}(k)}-\frac{Z^2}{k^2 g(\Lambda)}\right)\,. \label{eqA3}
\end{equation}
Moreover, following \eqref{defL}:
\begin{equation}
L_k(\eta)=\eta L_k^{(1)}+L_k^{(2)}\,,
\end{equation}
where the $L_k^{(j)}$ are independent of $\eta$; it is easy to check, from definition \eqref{defL} that $\mathcal{S}_{n}-\mathcal{S}_{n-1}=-L_n^{(1)}$. Therefore:
\begin{equation}
L_3+\eta \bar{A}_{3,0}=L_3^{(2)}+ \frac{1}{2}\left(\frac{1}{\bar{g}(k)}-\frac{Z^2}{k^2 g(\Lambda)}\right)\,,
\end{equation}
and the $\beta$-function $\beta_g$ becomes:
\begin{equation}
\beta_g=(d-1)\bar{g}+4\bar{g}^2L_3^{(2)}+\bar{g}^2\frac{2\eta Z^2}{k^2g(\Lambda)}\,.
\end{equation}
This is a non-tractable equation, depending on the initial conditions. For $\eta=0$, the difficulty to think about this equation disappears, and we recover the precedent result, with $\theta=d-1$. However, it has two simpler and interesting limit cases. In the deep UV limit $k \approx \Lambda$; we must have:
\begin{equation}
\bar{g}^2\frac{2\eta Z^2}{k^2g(\Lambda)} \approx 2\eta \bar{g}\,,
\end{equation}
so that the $\beta$-function reduces to:
\begin{equation}
\beta_g=((d-1)-2\eta)\bar{g}+4\bar{g}^2L_3^{(2)}\,.\label{deepUV}
\end{equation}
In contrast, let us consider the intermediate regime $\Lambda\gg k \gg 1$, far from the deep UV regime, but also far enough from the deep IR so that non-melonic terms are is neglected. Remembering that from the power counting $g(\Lambda) \sim \Lambda^{-2}$; we deduce that the $\eta$ dependent term dominates the flow, and
\begin{equation}
\beta_g \approx \bar{g}^2\frac{2\eta Z^2}{k^2g(\Lambda)}\,. \label{UV}
\end{equation}
Equation \eqref{UV} vanish only for $\bar{g}=0$. However, the first equation \eqref{deepUV} has a more interesting fixed point structure. Solving the equation $\eta=-2\bar{g}(\eta L_2^{(1)}+L_2^{(2)})$ as:
\begin{equation}
\eta= \frac{-2\bar{g} L_2^{(2)}}{1+2\bar{g} L_2^{(1)}}\,, \label{generaleta}
\end{equation}
we get for $\beta_g=0$ the condition:
\begin{equation}
\big((d-1)(1+2\bar{g} L_2^{(1)})+ 4\bar{g} L_2^{(2)}\big)+4\bar{g}(1+2\bar{g} L_2^{(1)})L_3^{(2)}=0
\end{equation}
where we assumed $\bar{g}\neq 0$. Numerical investigations, using the regulators used in this section show that the resulting fixed point match with the results of the previous section. In rank $3$, we recover a fixed point having essentially the same characteristics as the fixed point obtained from a quartic truncation. More interestingly is the behavior of this solution with the rank $d$ of the tensor . On Figure \ref{FigurePlotd}, we show the behavior of the critical exponent with the rank using the scheme $S_2$, and we show that $\theta \geq d-1$, and converge weakly toward this limit. \\
\begin{figure}
\includegraphics[scale=0.5]{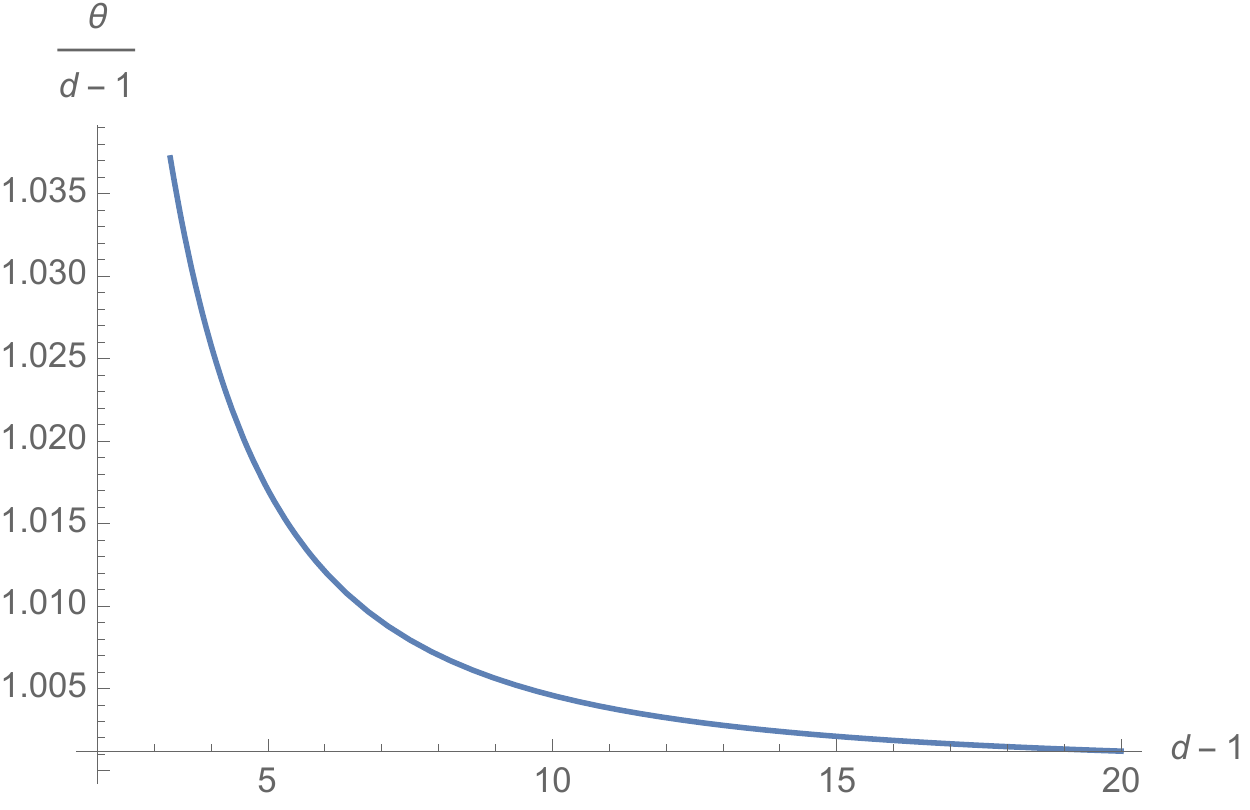}
\caption{Dependence of $\theta$ with the rank using the regularization scheme $S_2$.}
\end{figure} \label{FigurePlotd}
Now, let us consider the first condition $A_{3,0}=0$. From equation \eqref{eqA3},
\begin{equation}
-L_3^{(1)}+\frac{1}{2}\left(\frac{1}{\bar{g}(k)}-\frac{Z^2}{k^2 g(\Lambda)}\right)=0\,.
\end{equation}
Noting that $L_3^{(1)}$ is a pure number, this equation can be translated locally as a differential equation:
\begin{equation}
\dot{\bar{g}}=((d-1)-2\eta)\left(\frac{Z^2}{g(\Lambda) k^2}\right)\, \bar{g}^2\,.
\end{equation}
In the deep UV regime, it may be approximated by the most suggesting expression:
\begin{equation}
\dot{\bar{g}}\approx ((d-1)-2\eta)\, \bar{g}\,.
\end{equation}
The $\beta$-function behaves as there are no interaction at all. The only trace of the non-Gaussian measure is in the definition of the anomalous dimension and for this reason we refer to this solution as the \textit{purely scaling limit}. In addition to the Gaussian fixed point, we get the condition:
\begin{equation}
\eta^*=\frac{d-1}{2} \,\Rightarrow \, \bar{g}^*=-\frac{1}{2}\frac{d-1}{2d L_2^{(2)}+(d-1)L_2^{(1)}}\,;
\end{equation}
leading to the critical exponent:
\begin{equation}
\theta=+2\frac{\partial \eta}{\partial \bar{g}}\big\vert_{\bar{g}=\bar{g}^*}=(d-1)\left[1+\frac{L_2^{(1)}}{L_2^{(2)}}\frac{d-1}{2d}\right]\,.
\end{equation}
Once again, the numerical investigations based on the regulator considered in this paper show that this critical exponent is always bigger than $d-1$. Moreover, we showed that, except for the solution $\eta=0$, all the solutions of \eqref{compatibilityrel} do not allows to obtain autonomous local systems, without dependence on the initial conditions. To summarize:
\begin{claim} \label{claim}
In the UV regime, the compatibility with the melonic structure equations imposes that for any full ultra-local approximation of the effective action, involving only connected bubbles we must have $\eta=0$. Moreover, for the complete truncation, when all the graphs are took into account, we get $\theta_{\text{op}}=d-1$.
\end{claim}
This conclusion are obviously in accordance with our results of the previous subsections. In particular, we showed that any truncation which reduces the Ward identity violation, and therefore improve the reliability of the purely local truncation improves as well the rate of convergence toward the limit $\theta_{\text{op}}=d-1$. In the next subsection, we will consider the effect of disconnected pieces, from a ‘‘dressed" parametrization of the local theory space. \\

Even to close this section, we let us add another important remark. From the structure equations between $2n$, $4$ and $2$-point observables, we were able to close the infinite hierarchy of flow equations in the melonic sector. In this sense, these equations take into account the whole melonic sector. Interestingly, we do not find more than one, or eventually two interacting fixed points. This strongly contrast with the results obtained in the melonic sector in the previous sections, using large local truncations, where a large number of fixed points were found. Some of these fixed points were interpreted as an artefact of the truncation, but a certain number of them, with more than one relevant direction, seems to stay in high trunks. We see now that these fixed points are artefacts of the truncation as well, which does not take into account the strong relation coming from the structural equations. This appears to be a new effect of the pathology which can appear when the constraints inherent in a given sector are ignored, making all the more difficult confidence in the results coming from a local truncation. Finally, the reader may be wondering what happens when you impose the condition $\eta=0$ for the truncation. This can be easily checked, for instance using the scheme $S_2$. With the choice $\alpha=4/3$, we show from equation \eqref{etaS2} that $\eta$ vanish. This condition does not vanish $\mathcal{L}_2$, and the violation of the modified Ward identity holds. However, the result seems to be strongly improved in the light of the exact results obtained in this section. In particular, we show that for large melonic truncations, up to order $8$ taking into account all the melons and up to order $20$ for the non-branching sector, we find an interacting fixed point with one relevant direction; whose critical exponent is always exactly equal to $2$. Once again, this result goes in the direction of our conclusions, and it seems that by taking into account of the structural equations has comparatively greater importance even than the Ward identities concerning the convergence of the flow. However, the fixed point in question has a very bad characteristic, effective high valence couplings with very large values (of order $10^{100}$); meaning that the flow has moved away considerably from the Gaussian point, and once again, highlighting a very strong dependence on non-universal quantities at the choice of regularization.

\subsection{Closing hierarchy around the full quartic sector}

Let us briefly consider the influence of disconnected melonic pieces on the results of the previous section. As discussed above, the disconnected pieces appear as soon as $\nu\neq 1$. First of all, note that the exact relations as \eqref{effcoupling} and \eqref{sixkernel} hold, independently with the parametrization used in the phase space. Now we have the following important question which needs to be solved: What is the condition satisfied by this parametrization in agreement with the exact relation at the leading order sector? \\

The equation \eqref{betaexact} holds. However, the equation \eqref{betaEVE} have to be modified by the coupling that we called $h_3$ in section \ref{sectiondisco}:
\begin{align}
\nonumber \beta_{g,EVE}=&((d-1)-2\eta)\bar{g}(k)+4\bar{g}^2(k) L_{3}(\eta)\\
&\qquad -48\, \bar{A}_{3,0} \bar{g}^3 L_{2}(\eta)-\bar{h}_3 K_2(\eta) \,. \label{betaEVEdisco}
\end{align}
The compatibility with equation \eqref{effcoupling} therefore requires:
\begin{equation}
-48\, \bar{A}_{3,0} \bar{g}^3 L_{2}-\bar{h}_3 K_2=4\bar{g}^2\eta \bar{A}_{3,0}\,.
\end{equation}
Moreover, the expression for the anomalous dimension receives a contribution for the disconnected quartic coupling (we keep the notation $g_1$ used in section \ref{sectiondisco}):
\begin{equation}
\eta=-6\bar{g} L_2-2\bar{g}_1 K_2\,.
\end{equation}
Therefore, assuming $\eta\neq 0$, we get the relation:
\begin{equation}
8\bar{g}^2\bar{A}_{3,0}(-3\bar{g} L_2+\bar{g}_1 K_2)-\bar{h}_3 K_2=0\,.
\end{equation}
We then have the explanation of the phenomenon observed in the section \ref{sectiondisco}, i.e. the existence of the relations making the disconnected couplings dependent on the other couplings. From the previous section, we know that the presence of $\bar{A}_{3,0}$ introduces a spurious dependence on the initial condition. Moreover, a direct inspection show that, adjusting $h_3$ to compensate the term sharing the factor $\bar{A}_{3,0}$ ultimately requires $\eta=0$, which implies that:
\begin{equation}
3\bar{g} L_2+\bar{g}_1K_2=0\,,
\end{equation}
and:
\begin{equation}
48\bar{g}^3\bar{A}_{3,0}L_2+\bar{h}_3K_2=0\,,
\end{equation}
and then discards the two last terms of \eqref{betaEVEdisco}. These two equations moreover ensure that $\dot{\bar{g}}=0\Rightarrow \dot{\bar{g}}_1=\dot{\bar{h}}_3=0$. Therefore, the difficulties arising from the disconnected pieces seems to be solved. The strong relation between observables make them dependent on other couplings, explaining the apparent success of ultralocal truncation. Obviously, this reasoning remains fairly qualitative, and we will endeavor to remedy the shortcomings in our future work. However, at this stage, we can ask ourselves if the convergence problems notified in section \ref{sectiondisco} do not comes quite simply by taking into account these constraints.

\section{Optimization criteria}\label{section3}

At this stage we must specify our criteria for judging the quality of an approximation. Let us recall that an approximation is essentially the combination of two choices, the choice of a particular parameterization of the space of the phases given by $\Gamma_k$, and the choice of a regulator $r_k$. Usually, in FRG literature, optimization has a precise meaning. In the symmetric phase, all the loop integrals involved in the flow equations involves the effective propagator $P:=C^{-1}+r_k$, where $C$ denote the bare propagator of the theory. This effective propagator has a minimum, whose position depends on the choice of $r_k$. More generally, the development of effective action takes place around a non-zero vacuum, and the minimum of $ P $, we remove the risk of seeing a singularity develop around the non-zero vacuum. A regulator is then optimal when the lower bound of the free propagator $P$ is minimal. This is the sense given by Litim for optimization, and this is a very general criterion, essentially independent on the specificity of the problem that we consider. The so-called Litim regulator is optimal in this sense. This is as well the case of the regulator with $\alpha >1$ that we considered in schemes $S_1$ and $S_2$. However, in this paper, the notion of optimization is quite different and may be summarized as follows. Among a more and less large set of regulators (optimal in the Litim sense), the optimal choice(s) is such that:
\begin{itemize}
\item The calculation of universal quantities such that the critical exponents are as close as possible to the exact results available, or the speed of convergence to these exact results is most important when the order of the truncation increases.
\item The disagreement with the set of constraints on the observables (coming for example from the symmetries of field theory) remains as small as possible, and does not increase with the order of truncation, to the orders corresponding to the effects that we hope to update.
\item The computation of the universal quantities, in a scheme satisfying the two previous requirements, should not change too much under slight modifications of the regulator.
\end{itemize}
We showed in the previous section how the two first requirements work for practical calculations.

\section{Derivative expansion - a first look}\label{section4}
\begin{figure}
\includegraphics[scale=0.6]{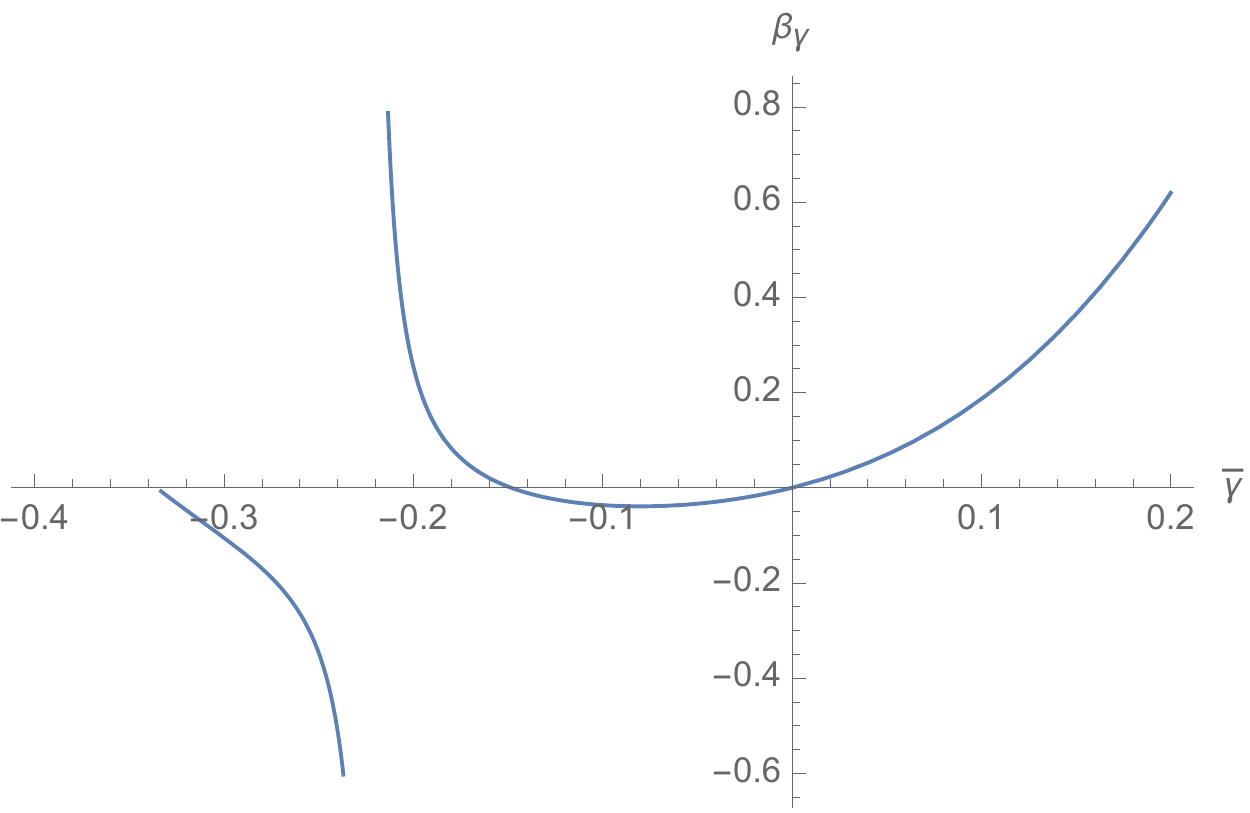}
\caption{Numerical plot of the function $\beta_\gamma$. We see that it has two zeros, for $\bar{\gamma}=0$ and $\bar{\gamma}\approx -0.15$. Moreover, it has a vanishing limit point at $\bar{\gamma}\approx -0.33$, beyond which the function becomes imaginary.}\label{figfloteffectif}
\end{figure}

In this section, we provide a first look about an alternative way to deal with modified Ward identities violations: introducing derivative operators on the truncation itself. We do not provide a deep investigation on this effect, the only interest is to compare this method with the other one considered in the previous section. We provide a solution at the same level of approximation, from an approximate solution solving only the first Ward identity \eqref{Ward1}.
We consider a local truncation of the form:
\begin{align}
\nonumber \Gamma_k[M,\bar{M}]&=\, \gamma(k)\left(\vcenter{\hbox{\includegraphics[scale=1]{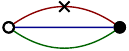} }}+\vcenter{\hbox{\includegraphics[scale=1]{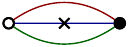} }}+\vcenter{\hbox{\includegraphics[scale=1]{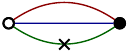} }}\right)\\
& +Z(k)\,\vcenter{\hbox{\includegraphics[scale=1]{melon0.pdf} }}+{g}(k)\,\sum_{i=1}^d \vcenter{\hbox{\includegraphics[scale=0.8]{melon4.pdf} }}+\cdots\,,
\end{align}
where the cross on the link of color $i$ denotes insertion of the ‘‘derivative" operator $n_i/k$. For instance:
\begin{equation}
\vcenter{\hbox{\includegraphics[scale=1]{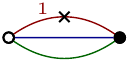} }} \equiv \sum_{n_1,n_2,n_3}\, \frac{n_1}{k} \bar{M}_{\vec{n}\,} M_{\vec{n}}\,.
\end{equation}
With this definition, we have (see equation \eqref{Ward1}):
\begin{equation}
\gamma(k)\equiv \frac{d}{dx_1} \,\gamma_k^{(2)}\,,
\end{equation}
and therefore, from \eqref{Ward1}, $\gamma(k)$ must be related to $\bar{g}(k)$ as:
\begin{equation}
2\bar{g}\,\bar{\mathcal{L}}_2=-\bar{\gamma}(k)\,. \label{ward1bar}
\end{equation}
We introduced the notation $\bar{\mathcal{L}}_n$, denoting the dimensionless version of the quantity $\mathcal{L}_n$, discarding the $k$ and $Z$ dependence. Moreover, we defined the renormalized $\bar{\gamma}$ as
\begin{equation}
\bar{\gamma}(k):=\frac{1}{Z}\gamma(k)\,.
\end{equation}
The flow equations for $g$ and $\gamma$ can be easily deduced from \eqref{eq1}. Indeed, the expression for $\beta_g$ remains the same as computed in \eqref{flowgeneral}:
\begin{equation}
\beta_g=((d-1)-2\eta)\bar{g}+4\bar{g}^2 L_3(\eta)-(3\bar{h}_1+2\bar{h}_2)L_2(\eta)\,,
\end{equation}
where $\eta$ is given by equation \eqref{generaleta}. The equation for $\gamma$ can be deduced taking the first derivative of \eqref{eve1} with respect to $n_1$. It is easy to check that the derivative of the effective loop involving $\dot{r}_k (G^{(2)})^2$ vanish in the large $k$ limit. Then only the derivative of the effective vertex contributes; we then get ($\beta_\gamma\equiv \dot{\bar{\gamma}}$):
\begin{equation}
\beta_\gamma= -\eta \bar{\gamma}-2\frac{d\bar{\pi}_k^{(2)}}{dx_1}(0,0)L_2(\eta)
\end{equation}
Equation for the first derivative of the effective $4$-point vertex can be obtained taking fourth order derivative of the full Ward identity \eqref{WI}, and vanishing external momenta. We can easily prove that:
\begin{equation}
\left(\left(6h_1+4h_2 \right) \mathcal{L}_2-8g^2 \mathcal{L}_3 \right)=-2\frac{d}{dn}\pi^{(2)}_k(0,0)\,. \label{WI20}
\end{equation}
Similarity from \eqref{WI20}, we obtain:
\begin{equation}
\beta_\gamma=-\eta \bar{\gamma}-2\left(\left(3\bar{h}_1+2\bar{h}_2 \right) \bar{\mathcal{L}}_2-4\bar{g}^2 \bar{\mathcal{L}}_3 \right) L_2(\eta)\,.
\end{equation}
We focus on Litim regulator, and $\bar{\mathcal{L}}_n$ can be easily computed. Using the notation of section \ref{quartic}, it is not hard to check that ($\alpha=1$):
\begin{equation}
\bar{\mathcal{L}}_n=d^2\, \int_0^1\frac{x^{n-1}}{(1+d\bar{\gamma} x^2)^2}\equiv \chi_{n-1,2}\,,
\end{equation}
where we introduced $\chi_{p,q}$ defined as:
\begin{equation}
\chi_{p,q}:=d^2\, \int_0^\alpha\frac{x^{p}}{(1+d\bar{\gamma} x^2)^q}\,.
\end{equation}
For $d=3$, the integral can be easily computed using a simple integration by part. For instance:
\begin{equation}
\bar{\mathcal{L}}_2= d^2\, \int_0^1\frac{x}{(1+d\bar{\gamma} x^2)^2}=-\frac{d^2}{2d \bar{\gamma}}\, \int_0^1 \frac{d}{dx}\, \frac{1}{1+d\bar{\gamma} x^2}\,,
\end{equation}
leading to:
\begin{equation}
\bar{\mathcal{L}}_2=-\frac{d^2}{2}\,\frac{1}{1+d\bar{\gamma}}\,. \label{L2exp}
\end{equation}
We will need also to the explicit expression for $\mathcal{L}_3$:
\begin{align}
\nonumber \bar{\mathcal{L}}_3&= d^2\, \int_0^1\frac{x^2}{(1+d\bar{\gamma} x^2)^2}\\\nonumber
&=-\frac{9}{6\bar{\gamma}}\left( \frac{x}{1+3\bar{\gamma} x^2}\bigg\vert_0^1-\int_0^1dx\, \frac{1}{1+3\bar{\gamma} x^2}\right)\\
&=-\frac{9}{6\bar{\gamma}}\left(\frac{1}{1+3\bar{\gamma}}-\frac{1}{\sqrt{3\bar{\gamma}}} \arctan(\sqrt{3\bar{\gamma}}) \right)
\end{align}
From \eqref{L2exp} the Ward identity becomes:
\begin{equation}
\bar{\gamma}=d^2\bar{g}\, \frac{1}{1+d\bar{\gamma}}\,, \label{gammaexplicit}
\end{equation}
which can be solved as\footnote{The other solution $\bar{\gamma}=-\frac{1}{6}\left(1+\sqrt{1+108 \bar{g}}\right)$ does not vanish for $\bar{g}\to 0$. }:
\begin{equation}
\bar{\gamma}=-\frac{1}{6}\left(1-\sqrt{1+108 \bar{g}}\right)\,.
\end{equation}
Then, differentiating this relation with respect to $t=\ln(k)$, we get:
\begin{equation}
\beta_\gamma=\frac{9}{\sqrt{1+108\bar{g}}} \beta_g=\frac{9}{1+6\bar{\gamma}}\beta_g\,.
\end{equation}
Then, from explicit expressions for $\beta_g$ and $\beta_\gamma$, we deduce a relation between $3\bar{h}_1+2\bar{h}_2 $ and $\bar{g}$ and $\bar{\gamma}$; $3\bar{h}_1+2\bar{h}_2=h(\bar{g},\bar{\gamma})$, with:
\begin{align}
h(\bar{g},\bar{\gamma})=(1+6\bar{\gamma})\frac{\eta A(\bar{\gamma})+18\bar{g}\frac{(1-\eta)+9\bar{g}\left(\frac{\eta}{10}+\frac{1}{2}\right)}{1+6\bar{\gamma}}}{{27}\left(\frac{\eta}{4}+1\right)\left(1+\frac{1+6\bar{\gamma}}{1+3\bar{\gamma}}\right)}\,,
\end{align}
and
\begin{equation}
A(\bar{\gamma}):=\bar{\gamma}-\frac{2}{3}\left(1-\frac{1+3\bar{\gamma}}{\sqrt{3\bar{\gamma}}} \arctan(\sqrt{3\bar{\gamma}})\right)\,.
\end{equation}
We can remark that the non-branching sector is relevant for fixed point investigations, especially for the double scaling limit. Vanishing $h_2$, and from the relation \eqref{gammaexplicit}, we have:
\begin{equation}
3\bar{h}_1=h\left(\frac{\bar{\gamma}(1+3\bar{\gamma})}{9},\bar{\gamma}\right)=:H(\bar{\gamma})\,,
\end{equation}
and the flow equations reduces to a single relation:
\begin{align}
\nonumber\beta_\gamma=&-\eta \bar{\gamma} +27\frac{H(\bar{\gamma})}{1+3\bar{\gamma}}\left(1+\frac{\eta}{4}\right)\\
&+\frac{\eta}{9}\left(1-\frac{1+3\bar{\gamma}}{\sqrt{3\bar{\gamma}}} \arctan(\sqrt{3\bar{\gamma}})\right)\,,\label{floteffectif}
\end{align}
where in this equation $\eta$ have to be expressed in term of $\bar{\gamma}$, explicitly:
\begin{equation}
\eta\equiv -\frac{4(1+3\bar{\gamma})\bar{\gamma}}{2+(1+3\bar{\gamma})\bar{\gamma}}\,.
\end{equation}
Interestingly, taking into account the first order deviation from ultralocality allows to describe all the non-branching sector with a single flow equation, equation \eqref{floteffectif}. Indeed, we expressed $h_1$ in terms of $\bar{g}$ and $\bar{\gamma}$; and the flow equation for $\bar{h}_1$ allows to fix $u_1$, the octic coupling, and so one. Obviously, we discarded all the higher derivatives, and the momentum dependence of the observables with valence higher than $2$. Nevertheless, we illustrate on this simple example how the inter-dependence between local and non-local observables coming from Ward identities may have consequence on entire sectors. \\
We may investigate the fixed point structure of the flow equation \eqref{floteffectif}. Figure \ref{figfloteffectif} represents the effective $\beta$-function $\beta_\gamma$. Among the zeros of the $\beta$-function, only t $\bar{\gamma}\approx -0.15$ seems to be relevant. The corresponding critical exponent is $\theta\approx 1.77$ and the anomalous dimension $\eta \approx 0.17$. As we observed, taking into account the first Ward identity this improves strongly the result. Note that our truncation being non-local, the bound $\theta_{\text{op}}=d-1$ for local truncations does not hold. \\

In regard to the EVE, as mentioned at the beginning of the section \ref{sectionEVE}, the fact that we take into account all the momentum dependence of the effective vertices allow in principle to go beyond the dressed local potential approximation that we considered. There are in particular a very interesting aspect or the melonic EVE: In the melonic sector, all the relations between effective vertex functions due to the EVE are compatible with Ward identities; meaning that no additional assumptions are required to deal with them. The same thing has been observed for a sub-leading sector in \cite{Lahoche:2018oeo}, and it is tempting to conjecture that it must be a general property of EVE, sector by sector in the $1/N$ expansion. This property can be easily checked for the melonic sector. From section \ref{sectionEVE}, we know that the melonic self energy per color $\sigma_k(n)$ must satisfy the closed equation \eqref{closed}. Taking the first derivative with respect to the external momenta $n$, we get (note that we set $\nu=d$ for this section):
\begin{equation}
\frac{d\sigma_k}{dn} = 2g\, \sum_{\vec{n}}\, \delta_{n_1n}\, \frac{- \frac{d\sigma_k}{dn} +\frac{dr_k}{dn} }{(1-\sum_{i=1}^d\sigma(n_i)+r_k(\vec{n}\,))^2}\,. \label{derivclosed}
\end{equation}
From the definitions of $\mathcal{L}_p$ (see equation \eqref{defLp}), $\mathcal{A}_{m,n}$ (see equation \eqref{defAp}) and $\gamma\equiv -d\sigma_k/dn(n=0)$; we get, setting $n=0$:
\begin{equation}
-\gamma=2g \mathcal{L}_2+2g\gamma \mathcal{A}_{2,0}\,,
\end{equation}
or simply:
\begin{equation}
\gamma=-2\left(\frac{g}{1+2g \mathcal{A}_{2,0}}\right) \mathcal{L}_2\,.
\end{equation}
Then, from equation \eqref{effcoupling}, the bracket term in nothing but $\pi^{(2)}_k(0,0)$, and the previous equation reduces to the Ward identity \eqref{Ward1}. The same compatibility can be checked for higher order Ward identities. For instance, restricting to the non-branching sector, we deduce from \eqref{WI20}:
\begin{equation}
\left(\pi^{(3)}_k(0,0,0) \mathcal{L}_2-8(\pi^{(2)}_k(0,0))^2 \mathcal{L}_3 \right)=-2\frac{d}{dn}\pi^{(2)}_k(0,0)\,. \label{WI2}
\end{equation}
in the same way as we derive in the equation \eqref{Ward1}. From equation \eqref{effcoupling}, the derivative can be easily computed, leading to:
\begin{equation}
\pi^{(3)}_k(0,0,0)\mathcal{L}_2=4(\pi^{(2)}_k(0,0))^2 \left[\frac{dA_{2,0}}{dn}+2\mathcal{L}_3\right]\,.
\end{equation}
Moreover, as for equation \eqref{derivclosed}, it is easy to compute the derivative of $A_{2,n}$. Using once again the fact that, in the melonic sector the free energy decomposes as $\Sigma_k(\vec{n}\,)=\sum_i \sigma_k(n_i)$, we get straightforwardly:
\begin{equation}
\frac{dA_{2,0}}{dn}=-2\sum_{\vec{n}}\, \delta_{n_10}\, \frac{- \frac{d\sigma_k}{dn} +\frac{dr_k}{dn} }{(1-\sum_{i=1}^d\sigma(n_i)+r_k(\vec{n}\,))^3}\,,
\end{equation}
and therefore:
\begin{equation}
\frac{dA_{2,0}}{dn}-2\mathcal{L}_3=-2\gamma A_{3,0}\,.
\end{equation}
Finally, from equation \eqref{effcoupling} and Ward identity \eqref{Ward1}, we get:
\begin{equation}
\pi^{(3)}_k(0,0,0)\mathcal{L}_2=2^4(\pi_k^{(2)}(0,0))^3 \gamma A_{3,0} \mathcal{L}_2\,,
\end{equation}
from which we recognize the expression of $\pi^{(3)}_k(0,0,0)$ given by EVE, equation \eqref{sixkernel}. Note that the proof seems to be very dependent on the fact that equation \eqref{melodecomp} hold, especially for $2$-point functions. Such a condition, however could be lost for sub-leading orders \cite{Lahoche:2018oeo}; which may request additional conditions regarding Ward's identities.

\section{Conclusion}\label{section6}

In this paper, we have essentially focused on the compatibility between local truncations and exact relations between observables at the large $N$ limit. These relations, moreover have different natures. The modified Ward identities come from the internal symmetry group used to define the allowed interactions, and the structure equations are nothing but ordinary Schwinger-Dyson equations in the melonic sector. With this respect, these relations may be understood in two different manners. On one hand, Schwinger-Dyson equations are the consequence of the formal Lebesgue measure involved in the path integral definition of the partition function \cite{Samary:2014tja}. One the second hand, the structure equations may be derived directly in the large $N$ limit as a consequence of the recursive definition of melons \cite{Lahoche:2018vun}. To put in a nutshell, we showed that:
\begin{itemize}
\item Accommodating these constraints, we improve the rate of convergence toward a given limit in a given sector, this limit depending on the sector that we consider.
\item The flow seems to be more sensitive to the structural constraints, arising from the Schwinger-Dyson equations, than to the symmetry constraint given by the modified Ward identities.
\end{itemize}
One expects that the second point is a consequence of the fact that, generating the flow requires a symmetry breaking, modifying the Ward identities, while the structure equations remain formally unchanged. Despite the existence of a fixed point having a single relevant direction, and reminiscent to the critical scenario of the double scaling, two difficulties appeared in the light of this study. The first is that a considerable number of fixed points generally accompany this; (fixed points which generally have more than one lifting direction, and can persist in high truncations). This is not clearly understood because, although it is tempting to interpret these fixed points as possible as a multi-critical fixed point, it corresponds to different limits beyond double scaling. It has been shown that an analysis taking into account the close relations between observables coming from the structural equations discard these residual fixed points. The effect of Ward's identities moreover seems less crucial. Indeed, it is easy to check that taking into account the locality constraint coming from the structure equations, $\eta=0$, provides a fixed point with one relevant direction having exactly the limit value $\theta = 2$ for any truncations (up to order 20). In contrast, to accommodate Ward identities, at least in the first order of derivative expansion implies slow convergence phenomena. These conclusions will be straightforwardly extended to the real models based on the internal group $\mathrm{O}(N)$, which probably remains in agreement with the conclusions of the recent work \cite{Lahoche:2019ocf} and which remains a subject of forthcoming work. \\

Thus, we then expect that this work provides a serious way of reflection and investigation on the methods used to compute the critical behavior of random tensor models. We have pointed out the crucial role played by exact functional relations, but we focused only on the leading order in the $1/N$ expansion. EVE method for sectors beyond melons quickly becomes intractable, as showed in \cite{Lahoche:2018oeo}. A promising way, outlined in this paper could be to "dress" a few complete sectors with truncations; taking care for each new magnitude explored for the coupling constants, the violation of the different exact relation is available. This strategy should be explored in an upcoming article and will help to increase confidence in the validity of the results made in the deep regions of the phase space. Finally, deeper investigations about integrability and regularity of the resulting RG flow has not to be done for RMM and RTM, for the considered approximations. These aspects will be considered in a work in progress.

\section{Acknowledgment}
V.L send special thanks to Laetitia Mercey for her constant support during the writing of this article.

\appendix

\section{Useful integrals}\label{Appendix}
Let us consider integrals of the form:
\begin{equation}
J=\int_0^\infty dx_1 dx_2\, \theta(\alpha-x_1-x_2) f(x_1+x_2)\,.
\end{equation}
Firstly, we set $x_i=y_i^2$, leading to:
\begin{equation}
J=4\int_0^\infty dy_1 dy_2 y_1 y_2\, \theta(\alpha-r^2) f(r^2)\,,
\end{equation}
where $r^2:=y_1^2+y_2^2$. Then, we use polar coordinates, $dy_1dy_2=rdr d\varphi$, $y_1=r \cos(\varphi)$, $y_2=r\sin(\varphi)$, leading to:
\begin{equation}
J=\left(\int_0^{\frac{\pi}{2}} d\varphi\, 2\sin(\varphi)\cos(\varphi)\right) \,\int_0^\alpha dr^2 r^2\theta(\alpha-r^2)f(r^2)\,.
\end{equation}
Where we restricted our analysis to the angular domain in the region where both $y_1$ and $y_2$ are positives. Finally:
\begin{equation}
J=\int_0^\alpha dx xf(x)\,.
\end{equation}
In the same way, we consider the three dimensional integral:
\begin{equation}
K=\int dx_1 dx_2 dx_3 f(x_1+x_2+x_3) \theta(\alpha-x_1-x_2-x_3)\,.
\end{equation}
We introduce $y_i$ defined as $x_i=y_i^2$, and $r^2=\sum_i y_i^2$, so that:
\begin{equation}
K=8\int d^3y (y_1y_2y_3) f(r^2) \theta(\alpha-r^2)\,.
\end{equation}
Then, introducing the polar coordinates:
\begin{align*}
y_1&=r\cos(\vartheta)\,,\\
y_2&=r\cos(\varphi)\sin(\vartheta)\,,\\
y_3&=r \sin(\varphi)\sin(\vartheta)\,.
\end{align*}
then the integral $K$ becomes:
\begin{align*}
K=&8\int_0^{\frac{\pi}{2}} d\vartheta \int_0^{\frac{\pi}{2}}d\varphi \cos(\vartheta)\sin^3(\vartheta)\cos(\varphi)\sin(\varphi)\\
&\times \int_0^{\sqrt{\alpha}} dr r^5 f(r^2)\,.
\end{align*}
The angular integrals can be easily computed, and we find:
\begin{equation}
8\int_0^{\frac{\pi}{2}} d\vartheta \int_0^{\frac{\pi}{2}}d\varphi \cos(\vartheta)\sin^3(\vartheta)\cos(\varphi)\sin(\varphi)=1\,,
\end{equation}
and finally, introducing $x=r^2$,
\begin{equation}
K=\frac{1}{2}\int_0^\alpha x^2f(x)dx\,.
\end{equation}


\onecolumngrid

\end{document}